\documentclass[11pt,twoside]{article}
\usepackage{atmp}
\usepackage{amsmath,amssymb}
\usepackage{epsfig}

\numberwithin{equation}{section}

\newcommand{\grad}[0]{\nabla\!}
\newcommand{\hateq}{\widehat{=}}

\newcommand{\Lie}[0]{{\cal L}}

\makeatletter
\newcommand{\pback}[1]{{
   \let\@rrow=\leftarrowfill
   \mathchoice{\AIN@stemPullBack{#1}{\@rrow}}{\AIN@stemPullBack{#1}{\@rrow}}
     {\AIN@indxPullBack{#1}{\@rrow}}{\AIN@indxPullBack{#1}{\@rrow}}}
   \vphantom{#1}}

\newcommand{\AIN@stemPullBack}[2]{
   \vtop{\mathsurround=0pt
   \ialign{##\crcr$\textstyle{#1}\strut$\crcr
     \noalign{\kern-0.4ex\nointerlineskip}{\tiny#2}\crcr}}}

\newcommand{\AIN@indxPullBack}[2]{
   \vtop{\mathsurround=0pt
   \ialign{##\crcr\hfil$\scriptstyle{#1}$\hfil\crcr
     \noalign{\kern+0.4ex\nointerlineskip}{\tiny#2}\crcr}}}
\makeatother

\newcommand{\man}{{\mathcal{M}}}

\newcommand{\dual}{{}^\star}

\newcommand{\half}[0]{\frac{1}{2}}

\let\puto=\overcirc

\def\kl{{\kappa_{(\ell)}\, }}
\def\t{{(t)}}

\def\l{\ell}
\def\ls{{(\l)}}

\def\ba{\begin{eqnarray}}
\def\ea{\end{eqnarray}}
\def\be{\begin{equation}}
\def\ee{\end{equation}}
\def\={\hateq}
\def\puto#1{\rlap{\raise.5ex\hbox{\char'27}}{#1}}
\def\D{{\cal D}}

\newcommand  {\Rbar} {{\mbox{\rm$\mbox{I}\!\mbox{R}$}}}
\def\hstar{\!\stackrel{\star}{}\!}

\newcommand{\Ricci}[2]{R_{ab}#1^a#2^b}

\url{gr-qc/0206024 }     

\begin{document}
\title{Isolated Horizons in \boldmath$2+1$ Gravity}
    \author{Abhay Ashtekar}
    \address{Center for Gravitational Physics and Geometry, Department
    of Physics, The Pennsylvania State University, University Park, PA
    16802, USA}
    \addressemail{ashtekar@gravity.phys.psu.edu}
    \author{Olaf Dreyer}
    \address{Perimeter Institute for Theoretical Physics, 35 King
    Street North, Waterloo, Ontario N2J 2W9, Canada}
    \addressemail{odreyer@perimeterinstitute.ca}
    \author{Jacek Wi\'sniewski}
    \address{Center for Gravitational Physics and Geometry, Department
    of Physics, The Pennsylvania State University, University Park, PA
    16802, USA}
    \addressemail{jacek@phys.psu.edu}

\markboth{\it Isolated Horizons in $2+1$ Gravity}{\it A. Ashtekar
et al.}

\vfill

\newpage

\begin{abstract}
Using ideas employed in higher dimensional gravity, non-expanding,
weakly isolated and isolated horizons are introduced and analyzed
in 2+1 dimensions. While the basic definitions can be taken over
directly from higher dimensions, their consequences are somewhat
different because of the peculiarities associated with 2+1
dimensions. Nonetheless, as in higher dimensions, we are able to:
i) analyze the horizon geometry in detail; ii) introduce the
notions of mass, charge and angular momentum of isolated horizons
using geometric methods; and, iii) generalize the zeroth and the
first laws of black hole mechanics. The Hamiltonian methods also
provide, for the first time,  expressions of total angular
momentum and mass of charged, rotating black holes and their
relation to the analogous quantities defined at the horizon. We
also construct the analog of the Newman-Penrose framework in 2+1
dimensions which should be useful in a wide variety of problems in
2+1 dimensional gravity.
\end{abstract}

\section{Introduction}
\label{s1} The zeroth and first laws of black hole mechanics apply
to equilibrium situations and small departures therefrom.  In
standard formulations of these laws, black holes in equilibrium
are represented by stationary space-times with regular event
horizons (see, e.g., \cite{review,sc}).  While this idealization
is a natural starting point, from a physical perspective it seems
quite restrictive.  (See \cite{ack,abf} for a detailed
discussion.)  To overcome this limitation, a new model for a black
hole in equilibrium was recently introduced for 3+1 (and higher)
dimensional gravity \cite{ack,abf,afk,abl2}. The generalization is
two-fold. First, one replaces the notion of an event horizon with
that of an isolated horizon.  While the former are defined only
retroactively using the fully evolved space-time geometry, the
latter are defined quasi-locally by suitably constraining the
geometry of the horizon surface itself. Second, one drops the
requirement that the space-time be stationary and asks only that
the horizon be isolated.  That is, the requirement that the black
hole be in equilibrium is incorporated by demanding only that no
matter or radiation fall through the horizon although the exterior
space-time region may well admit radiation. Consequently, the
generalization in the class of allowed space-times is enormous. In
particular, space-times admitting isolated horizons need not
possess \textit{any} Killing vector field; although event horizons
of stationary black holes are isolated horizons, they are a very
special case.  A recent series of papers \cite{abf,afk,abl2,ib}
has generalized the laws of black hole mechanics to this broader
context. The notion of isolated horizons has proved to be useful
also in other contexts in 3+1 dimensions, ranging from numerical
relativity to background independent quantum gravity: i) it plays
a key role in an ongoing program for extracting physics from
numerical simulations of black hole mergers
\cite{prl,dkss,abl1,abl3}; ii) it has led to the introduction
\cite{afk,cs,acs} of a physical model of hairy black holes,
systematizing a large body of results on properties of these black
holes which has accumulated from a mixture of analytical and
numerical investigations; and, iii) it serves as a point of
departure for statistical mechanical entropy calculations in which
all black holes (extremal or not) \textit{and} cosmological
horizons are incorporated in a single stroke \cite{ack,abck,abk}.

In recent years, 2+1-dimensional stationary black holes have drawn
a great deal of attention as simplified models for analyzing
conceptual issues surrounding black hole thermodynamics (see,
e.g., \cite{sc}). It is therefore natural to ask if the isolated
horizon framework can be constructed and used to extend the
standard 2+1-dimensional treatments. The purpose of this paper is
to provide such a framework, analyze the resulting horizon
geometry in detail and use it to generalize the zeroth and first
laws of black hole mechanics.

In Section \ref{s2}, we introduce the definitions of non-expanding
and weakly isolated horizons and derive their main consequences,
including the generalized zeroth law. While the basic definitions
are the same as in higher dimensions, some of their consequences
are different because of special features of the 3-dimensional
Riemannian geometry \cite{deser}. In particular, the Weyl tensor,
which plays an important role in higher dimensions, now vanishes
identically. Similarly, since in 2+1 dimensions black holes exist
only if the cosmological constant is non-zero, there is now an
inherent length scale in the problem. However, the spirit of the
analysis is the same as in higher dimensions: We extract, from the
notion of Killing horizons, the minimal structure that is needed
to generalize the laws of black hole mechanics. As in 3+1
dimensions, some of the structure becomes more transparent in
terms of null-triads. Therefore, in Appendix A we construct the
2+1-dimensional analog of the Newman-Penrose framework \cite{np}
and use it to elucidate the meaning and consequences of our
horizon boundary conditions.

In Section \ref{s3} we introduce the action principle, and in
Section \ref{s4}, the covariant phase space and the associated
Hamiltonian framework. While the overall procedure is the same as
in higher dimensions \cite{afk,abl2}, there is a significant
technical complication in the choice of boundary conditions at
infinity. In particular, while the electromagnetic potential falls
off as $1/r^{n}$ in $n+2$ spatial dimensions for $n>0$, one must
now allow it to \textit{blow up} logarithmically.  Since the
treatment of these boundary conditions is perhaps the most
difficult technical part of our analysis, they are spelled out in
detail separately in Appendix B. The end result is that there do
exist boundary conditions which suffice to make the action
principle, the symplectic structure and the Hamiltonians
well-defined. Using these structures, we establish the generalized
first law. As in higher dimensions, it arises as a consistency
condition for the evolution to be generated by a Hamiltonian;
there is thus an infinite family of first laws, one for each
time-like vector field in space-time, the evolution along which is
Hamiltonian.

In Section \ref{s5}, we consider the issue of introducing a
canonical notion of horizon energy which can be interpreted as
horizon mass. In the Hamiltonian framework, this problem reduces
to that of selecting (for each point in the phase space) a
preferred time-evolution vector field at the horizon. One expects
this choice to vary from one point in the phase space to another;
in the non-rotating case, one would expect the preferred vector
field to point along the null normal to the horizon, while in the
rotating case, one would expect it to have a non-zero component
also along the space-like, rotational direction at the horizon. As
in 3+1 dimensions, we resolve this problem by making use of known
stationary solutions \cite{btz,bdk,gc}.

Sections \ref{s3}-\ref{s5} focus on the infinite dimensional space
of histories and the phase space in presence of weakly isolated
horizons. In Section \ref{s6}, by contrast, we consider individual
space-times and analyze the geometrical structures and their
interplay with field equations at the horizon. Specifically, we
first show that non expanding horizons admit a natural derivative
operator $\D$ and study the geometrical information it encodes,
beyond the natural degenerate metric, and then use field equations
to isolate the freely specifiable parts of $\D$ on weakly isolated
horizons. As in higher dimensions \cite{afk} we introduce the
notion of isolated horizons using the derivative operator $\D$. In
contrast to higher dimensions, every non-expanding horizon can be
equipped with an isolated horizon structure simply by selecting an
appropriate null normal and, generically, this can be achieved in
a unique fashion. Readers who are primarily interested in the
notions of mass and angular momentum and black hole mechanics can
skip this section. Reciprocally, readers who are primarily
interested in the horizon geometry and field equations can go
directly to Section \ref{s6} after Section \ref{s2} without loss
of continuity. Section \ref{s7} summarizes the main results and
points out a subtlety in the definition of mass which arises again
because in 2+1 dimensions, the electromagnetic potential must be
allowed to diverge logarithmically at infinity.

Throughout this paper, we will set $8\pi G = c =1$. Since Newton's
constant has dimensions of inverse mass in 2+1 dimensions, now
mass and charge are dimensionless while angular momentum has
dimensions of length. As a general rule, arguments and proofs
which are parallel to those in higher dimensions
\cite{afk,abl2,abl1} are only sketched and differences are
emphasized.

\section{Definitions and geometrical structures}
\label{s2} In this Section we will define \textit{weakly isolated
horizons} and analyze their geometric properties. It is convenient
to proceed in two steps since certain preliminary results are
needed to state the final definition.

Let $\man$ be a three dimensional manifold with metric tensor
$g_{ab}$ of signature $(-\,+\,+)$. For simplicity, we will assume
that all manifolds and fields are smooth. Let $\Delta$ be a null
hypersurface in $(\man, g_{ab})$. A \textit{future directed} null
normal to $\Delta$ will be denoted by $\l$. The \textit{expansion}
$\theta_{(\l)}$ of $\l$ is defined by $\theta_{(\l)} \=
m^am^b\nabla_a \l_b$, where $\nabla$ is the derivative operator on
$(\man,g_{ab})$ and $m^a$ is any unit, space-like vector field
tangent to $\Delta$.%
\footnote{Throughout this paper, $\hateq$ will denote equality
restricted to the null surface $\Delta$.}
It is easy to check that the expansion is insensitive to the
choice of $m^a$. However, as the notation suggests, it does depend
on the choice of the null normal $\ell$; if $\l^\prime \= f\l$,
then $\theta_{(\l^\prime)} \= f\theta_{(\l)}$.

\subsection{Non-expanding horizons}
\label{s2.1}

\noindent\textbf{Definition 1}: A $2$-dimensional sub-manifold
$\Delta$ of a space-time $(\man,g_{ab})$ is said to be a
\textit{non-expanding horizon} if it satisfies the following
conditions:\\

(i) $\Delta$ is topologically $S^1 \times \Rbar$ and null;\\

(ii) The expansion $\theta_{(\l)}$ of $\l$ vanishes on $\Delta$
for any null normal $\l$;\\

(iii) All equations of motion hold at $\Delta$ and the
stress-energy tensor $T_{ab}$ of matter fields at $\Delta$ is such
that $-T^{a}_{\ b}\l^{b}$ is future directed and
causal for any future directed null normal $\l$.\\

Note that if conditions (ii) and (iii) hold for one null normal
$\l$ they hold for all.

The role of these conditions is as follows. The first condition
just ensures that the cross-sections of $\Delta$ are compact which
will in turn ensure that various integrals ---defining, e.g., the
symplectic structure and various Hamiltonians--- over these
cross-sections are well-defined. The second condition is the
crucial one. It directly implies that all horizon cross-sections
have the same length, which, following the terminology in the
literature \cite{sc} we will call the `horizon area' and denote by
$a_\Delta$.  It will also be convenient to introduce the notion of
the \textit{horizon radius} $R_\Delta$, defined by $a_\Delta =
2\pi R_\Delta$. Finally, as we will see below, condition (ii) also
implies that there is no flux of matter-energy across the horizon
and thus captures the intuitive notion that the black hole is
isolated. The last condition, (iii), is analogous to the dynamical
conditions one imposes at spatial infinity. While at infinity one
requires that the metric (and other fields) approach a specific
solution to the field equations (namely, `the classical vacuum'),
at the horizon one only asks that the field equations be
satisfied. The energy condition involved is very weak; it follows
from the much stronger dominant energy condition normally imposed.
All these conditions are satisfied on any Killing horizon (with a
$S^1$ cross-section) if gravity is coupled to physically
reasonable matter (including perfect fluids, Klein-Gordon fields,
Maxwell fields possibly with dilatonic coupling and Yang-Mills
fields).

We will now present three examples of non-expanding horizons:

\textsl{Example 1:} The paradigmatic example of a non-expanding
horizon in 2+1 dimensions is provided by the BTZ black
holes\cite{btz}. We begin by showing that the horizons of these
space-times trivially satisfy our Definition 1.

In Eddington-Finkelstein-like coordinates, the space-time metrics
of these black holes are given by
\begin{equation}\label{eqn:defefc}
  ds^2 = -(N)^2dv^2 + 2dv dr + r^2\left(d\phi +
  N^\phi dv\right)^2,
\end{equation}
where
\begin{equation}\label{eqn:nphi}
  N = \left(f(r) + \frac{J^2}{4\pi^{2}r^2}\right)^{1/2} \mbox{ and\ } N^\phi
=
  -\frac{J}{2\pi r^2},
\end{equation}
and
\begin{equation} \label{eqn:f}
    f(r) = -\frac{M}{\pi} + \frac{r^2}{l^2},
\end{equation}
the length $l$ being related to the cosmological constant
$\Lambda$ through
\begin{equation}\label{eqn:ccandl}
  \Lambda = -\frac{1}{l^2}.
\end{equation}
Thus, for any value $\Lambda$ of the cosmological constant, there
is a 2-parameter family of BTZ metrics, labeled by $M$ and $J$.
The metric coefficient $N$ vanishes at $r=r_\pm$, where
\begin{equation}\label{eqn:rpm}
  r^{2}_\pm = \frac{M l^2}{2\pi}\left\{ 1 \pm \left[ 1 - \left(\frac{J}{M
  l}\right)^2\right]^{1/2}\right\}\, .
\end{equation}
The 2-surface $r=r_+$ is the horizon of interest to us. Sometimes
it is convenient to have the mass $M$ and the angular momentum $J$
expressed in terms of $r_+$ and $r_-$:
\begin{equation}\label{eqn:MJr}
  M = \pi \frac{r_+^2 + r_-^2}{l^2}\ \ \ \ \ J = \frac{2\pi r_+ r_-}{l}\, .
\end{equation}
The surface $r= r_+$ is null with normal
\begin{equation}\label{eqn:deflbtz}
  \l = \partial_v - N^\phi(r_+)\partial_\phi.
\end{equation}
Since it is coordinatized by $v, \phi$, it has the required
topology, $S^1\times \Rbar$.  Since $\l$ is a restriction to the
horizon of a space-time Killing field $\partial_v -
N^\phi(r_+)\partial_\phi$ it follows that $\theta_{(\l)}$
vanishes. Finally, the third condition in the definition is
trivially satisfied because BTZ metrics are  vacuum solutions to
Einstein's equation.


\textsl{Example 2:} The first two conditions in our definition are
satisfied by the more general class of metrics (\ref{eqn:defefc}),
without the restriction (\ref{eqn:f}) on the form of the function
$f(r)$. If the function $f(r)$ is chosen to satisfy the weak
condition $(\partial_r f)\mid_{r= r_+} \le (r_+/l^2) $, the third
condition in the definition is also satisfied. Thus, we have a
very large class of \textit{generalized BTZ metrics} which admit a
non-expanding horizon. This class includes, in particular, the
metrics introduced in \cite{bdk}.

\textsl{Example 3:} Our final example is the charged, rotating
black hole solution first discovered by Cl\'{e}ment \cite{gc}. (It
was later independently found by Martinez, Teitelboim and Zanelli
(MTZ) \cite{mtz}, who also analyzed its physical properties.) It
is again a stationary axi-symmetric solution and is expressed in
terms of three parameters, $\bar{r}_0, \omega, Q$. As shown in
section \ref{s5.2}, they can be traded for mass $M$, angular
momentum $J$ and charge $Q$. However, as in higher dimensional
dilatonic black holes, the dependence of $M$ and $J$ on the
parameters appearing explicitly in the solution is quite
complicated (see Section \ref{s5}). Furthermore, in this case, the
electro-magnetic fields and the metric coefficients
\textit{diverge} (logarithmically) at infinity. Hence the very
meaning of mass and angular momentum is not a priori transparent.
Finally, if one simply sets $Q=0$, one obtains the BTZ metric with
$M=0$ and $J=0$; to obtain the non-trivial solutions in the BTZ
family, a more subtle limit has to be taken.

Cl\'{e}ment gave the metric in the form:
  \begin{equation}
   {\rm d}s^{2} = - N^{2} {\rm d}t^{2} + K^{2} ({\rm d}\phi +
   N^{\phi} {\rm d}t)^{2} + \frac{r^{2}}{K^{2}} \frac{{\rm
   d}r^{2}}{N^{2}}, \label{Cle}
  \end{equation}
where the functions $N,N^{\phi}$ and $K$ are given by
  \begin{eqnarray}
   N^{2} & = & \frac{r^{2}}{K^{2}} \left( \frac{r^{2}}{l^{2}} -
   \frac{\bar{l}^{2}}{2\pi l^{2}} Q^{2} \ln(r/\bar{r}_{0}) \right),
   \\
   N^{\phi} & = & - \frac{\omega}{2\pi K^{2}} Q^{2} \ln(r/\bar{r}_{0}),
   \\
   K^{2} & = & r^{2} + \frac{1}{2\pi} \omega^{2} Q^{2} \ln(r/\bar{r}_{0}),
\\
   \bar{l}^{2} & = & l^{2} - \omega^{2}
  \end{eqnarray}
The Maxwell field is given by:
  \begin{equation}
   {\mathbf F} = \frac{Q}{r} {\rm d}r \wedge ( {\rm d}t - \omega {\rm
   d}\phi).
  \end{equation}
In the Eddington-Finkelstein-like coordinates the metric becomes
   \begin{equation}
    {\rm d}s^{2} = -N^{2} {\rm d}v^{2} + \frac{2r}{K} {\rm d}v
    {\rm d}r + K^{2} \left( {\rm d}\phi + N^{\phi} {\rm d}v
    \right)^{2}. \label{EFm}
   \end{equation}
(As is usual in the passage to the Eddington-Finkelstein type
coordinates in the \textit{stationary} context,  the angle $\phi$
in ($\ref{EFm}$) is not the same as the one in ($\ref{Cle}$). In
the analysis of the horizon structure, we will use (\ref{EFm}).)
It is straightforward to check that the 2-surfaces $N=0$,
co-ordinatized by $v,\phi$, are non-expanding horizons.
\medskip

Although the conditions imposed in Definition 1 seem rather weak,
they have a number of interesting consequences. To explore them it
is often convenient to introduce, as in the Newman-Penrose
framework, a triad consisting of vectors $\l^a, n^a,$ and $m^a$ in
the neighborhood of the horizon $\Delta$. The vectors $\l^a$ and
$n^a$ are null and $m^a$, space-like. We choose $\l^a$ to be a
future pointing null normal of the horizon and then normalize
$n^a$ by requiring $\l^an_a \= -1$ and $m^a$ by requiring $m^am_a
= 1$. All other contractions vanish. (Thus, in contrast to the 3+1
dimensional NP framework, $m^a$ is now real and space-like rather
than complex and null.) On the horizon we further require $m^a$ to
be tangential to the horizon. Given such a triad, we can introduce
NP-like coefficients as in 3+1 dimensions. Appendix
\ref{sec:NP2plus1} gives the corresponding definitions and a
summary of important relations for these coefficients. It is often
convenient to use the triad so that the pull-back to $\Delta$ of
the 1-form $n$ is orthogonal to $S^1$ cross-sections of $\Delta$,
i.e., $\pback{dn} \= 0$ (so that, in the NP-like
framework of Appendix C, $\alpha \= \pi$).%
\footnote{Throughout this paper, an under-arrow will denote
pull-back.}
We will explicitly specify when this restriction is made.

We will conclude this subsection with a brief discussion of
geometric structures available on non-expanding horizons.

\medskip\noindent {(a)} \textit{Intrinsic metric of
$\Delta$}: Denote by $q_{ab}$ the pull-back of the space-time
metric $g_{ab}$ to $\Delta$; $q_{ab} \= g_{\pback{ab}}$. Since
$\Delta$ is a 2-dimensional, null sub-manifold of $(\man,
g_{ab})$, and $\l$ a null-normal to it, it follows that
 \begin{equation}
 q_{ab} \l^{b} \= 0; \quad \quad q_{ab} \= \underline{m}_{a}
 \underline{m}_{b}
 \end{equation}
for a unique 1-form $\underline{m}_a$, defined
\textit{intrinsically} on $\Delta$. Furthermore, as the explicit
calculation of spin coefficients in Appendix \ref{sec:NP2plus1}
shows, $d \underline{m} \= 0$. We will choose our NP triad such
that $m_{\pback{a}} = \underline{m}_a$.

\medskip\noindent {(b)} \textit{Properties of} $\l$: Since $\l^{a}$
is a null normal to $\Delta$, it is automatically twist-free and
geodesic. We will denote the \textit{acceleration} of $\l^a$ by
$\kappa_{(\l)}$
\begin{equation} \label{kappa}
\l^{a}\grad_{a}\l^{b} \= \kappa_{(\l)}\l^{b} \, .
\end{equation}
Note that the acceleration is a property not of the horizon
$\Delta$ itself, but of a specific null normal to it: if we
replace $\l$ by $\l^\prime \= f \l$, then the acceleration changes
via
\begin{equation}
 \kappa_{(\l^\prime)} \= f\kappa_{(\l)} + {\cal L}_\l f .
\end{equation}
(In the NP-type notation of Appendix A, $\kappa_{(\l)}$ is denoted
by $\epsilon$.)

\medskip\noindent {(c)} \textit{A natural connection 1-form on}
$(\Delta, \l)$: Since the expansion $\theta_{(\l)}$ (or, in the
framework of Appendix A, the NP-type coefficient $\rho$) vanishes,
and since in 2+1 dimensions there is no analog of the 3+1
dimensional shear, we conclude that given any vector field $X^a$
tangential to $\Delta$, we have:
\[ X^a \nabla_a \l^b \= X^a \omega_a \,\l^b  \]
for some ($\l$-dependent) 1-form $\omega_a$ on $\Delta$. In
particular, we have $\kappa_{(\l)} = \l^a \omega_a$. Thus, there
exists a one-form $\omega_{a}$ intrinsic to $\Delta$ such that
\begin{equation}\label{eqn:omegadefn}
\grad_{\!\pback{a}}\l^{b} \= \omega_{a}\l^{b}\, .
\end{equation}
$\omega_a$ will play an important role in this paper. (In the
NP-type framework of Appendix A, $\omega$ can be expressed in
terms of spin coefficients:
$\omega_a = \alpha \underline{m}_a - \epsilon n_a \equiv \alpha
\underline{m}_a - \kappa_{(\l)} n_a\,$.)
Under the rescaling $\l\rightarrow f \l$, the 1-form $\omega$
transforms as a connection:
\begin{equation}
{\label{eqn:omegachange}} \omega_a \rightarrow \omega_a +
\nabla_{\! \pback{a}} \ln f .
\end{equation}
A particular consequence of (\ref{eqn:omegadefn}) is:
\[
{\Lie}_\l\, q_{ab} \= 2\pback{\grad_{a}\l_{b}} \= 0\, ;
\]
every null normal $\l$ to $\Delta$ is a `Killing field' of the
degenerate metric on $\Delta$. Thus, our key condition in the
Definition  ---that $\l$ be expansion-free--- implies that
non-expanding horizons are Killing horizons of the
\textit{intrinsic geometry} to `first order'.

\textsl{Example 1:} What is the expression of $\omega_a$ in the
case of the BTZ black hole? On the horizon, let us choose the
triad vector $m^a$ as $m \= (1/r_+) \partial_\phi$. Then, a direct
calculation yields:
\begin{equation}\label{eqn:expromega}
    \omega_a = N^\phi m_a - \kappa_{(\ell)} n_a
\end{equation}
where the acceleration $\kappa_{(\l)}$ is given by
\begin{equation}\label{eqn:kappaBTZ}
  \kappa_{(\l)} = \frac{r_+}{l^2} - r(N^\phi)^2.
\end{equation}
(See the discussion of this example in appendix
\ref{sec:NP2plus1}.) As in higher dimensions \cite{abf,abl2}, the
angular momentum information is contained in the spatial component
of $\omega$.

\medskip\noindent {(d)}\textit{Conditions on the Ricci tensor}:
As in higher dimensions \cite{afk}, we can use the Raychaudhuri
equation to obtain conditions satisfied by the 3-dimensional Ricci
tensor at the horizon. Thus, by calculating $\Lie_{\l}\,\,
\theta_{(\l)}$ for a general null congruence $\l$ in terms of the
derivatives of $\l$ and the Ricci tensor and applying it to any
normal of a non-expanding horizon, we obtain:
\begin{equation}\label{eqn:rllzero}
  R_{ab} \l^a \l^b  \= 0.
\end{equation}
(For a derivation in the NP-type framework, see equation
(\ref{eqn:rll}) in appendix \ref{sec:NP2plus1}.) Next, let us use
the energy condition required in the Definition: $ P^a : \=
-T^a_{b}\l^b$ is future pointing, and time-like or null on
$\Delta$. Using the field equations
\begin{equation}\label{eq:einst}
R_{ab} - \frac{1}{2} R\, g_{ab} + g_{ab}\Lambda = T_{ab}
\end{equation}
and (\ref{eqn:rllzero}), we obtain $P_a\l^a \= 0$, whence, at the
horizon, $P^a$ is of the form $P^a \= f \l^a + g  m^a$. The energy
condition now implies $g \= 0$, i.e., the component $T^a_{\
b}\l^b$ is proportional to $\l^a$. The field equations then imply:
\begin{equation} \label{eqn:rlm} R_{ab}\l^a m^b \= 0.
\end{equation}

This constraint on the Ricci curvature has an important
consequence. Using the expression of the 3-dimensional Riemann
tensor in terms of the Ricci tensor in the equality $2 \nabla_{[a}
\nabla_{b]} \l^d \= R_{abc}{}^d \l^c$, and (\ref{eqn:omegadefn}),
it is straightforward to express $d\omega$ in terms of $R_{ab}\l^a
m^b$. (\ref{eqn:rlm}) now implies that $\omega$ is exact:
\begin{equation}  d\omega \= 0 .\end{equation}
(In the NP-type framework of Appendix \ref{sec:NP2plus1}, $R_{ab}
m^a \l^b$ can be expressed in terms of spin-coefficients as
$R_{ab} m^a \l^b \= \Lie_{\l}\, \pi - \Lie_{m}\, \kappa_{NP}$ and
$ d\omega \, \= $ $(\Lie_{\l}\, \pi - \Lie_m \, \kappa_{NP} )\,
m\wedge n$, whence (\ref{eqn:rlm}) implies that $\omega$ is
exact.) By contrast, in 3+1 dimensions, $d\omega$ is essentially
determined by the imaginary part of the Weyl tensor component
$\Psi_2$, which encodes the angular momentum information
\cite{abl2}. We will see that angular momentum information
continues to reside in $\omega$; it is just that, since the Weyl
tensor vanishes identically in 3 dimensions, we can no longer
further simplify that expression and rewrite it in terms of the
$\Psi_2$.

\medskip\noindent {(e)} \textit{Projective space}: Since $\l$ Lie-drags
the intrinsic metric $q_{ab}$ of $\Delta$, it is natural to pass
to the space $\hat{\Delta}$ of orbits of $\l$. We will conclude
the discussion of non-expanding horizons with a discussion of
$\hat{\Delta}$.

It follows from our topological restriction in Definition 1 that
$\hat\Delta$ has the topology of $S^1$. Denote by $\hat{\Pi}$ the
canonical projection map from $\Delta$ to $\hat\Delta$. Then,
since $q_{ab} \l^b \= 0$ and $\Lie_{\l}\, q_{ab} \= 0$, it follows
that there exists a metric $\hat{q}_{ab}$ on $\hat\Delta$ such
that ${q}_{ab} \= \hat{\Pi}_\star \hat{q}_{ab}$. The metric
$\hat{q}_{ab}$ on $\hat\Delta$ can be uniquely expressed as
$\hat{q}_{ab} = \hat{m}_a \hat{m}_b$ and $\underline{m}_a =
\hat{\Pi}_\star \hat{m}_a$.

\subsection{Weakly isolated horizons}
\label{s2.2}

Although non-expanding horizons already have a rather rich
structure, the notion is not sufficiently strong to be directly
useful to black hole mechanics. In particular, as we have seen,
there is a freedom to rescale the null normal via $\l^a
\rightarrow {\l^\prime}^a = f \l^a$ for any positive function $f$
on $\Delta$ under which the acceleration of $\l$ transforms via
$\kappa_{(\l^\prime)} = f \kappa_{(\ell)} + \Lie_{\l}\, f$.
Because of this rescaling freedom, $\kl$ will not be constant for
a generic choice of $\l$. Thus, on a general non expanding
horizon, we can not hope to establish the zeroth law. In this
sub-section, we will introduce a stronger definition by adding the
minimal requirements needed for a natural generalization of black
hole mechanics.

Let us begin by introducing an equivalence relation on the space
of null normals to a non-expanding horizon $\Delta$. The
transformation property (\ref{eqn:omegachange}) of $\omega_a$
under rescalings of $\l^a$ shows that $\omega_a$ remains unaltered
if and only if  $\l^a$ is rescaled by a constant. Therefore it is
natural to regard two null normals as equivalent if they differ
only by a (positive) constant rescaling. We will denote each of
these equivalence classes by $[\l]$. In what follows we will be
interested in non-expanding horizons $\Delta$, \textit{equipped
with} such an equivalence class $[\l]$ of null normals.

\medskip
\noindent \textbf{Definition 2}: A \textit{weakly isolated
horizon} $(\Delta, [\l])$ consists of a non-expanding horizon
$\Delta$, equipped with an equivalence class $[\l]$ of null
normals satisfying
\begin{equation} \label{i} {\cal L}_{\l} \omega \=0\,\,\, {\hbox{\rm for
all}}\,\, \l \in [\l].\end{equation}
\noindent As pointed out above, if this last equation holds for
one $\l$, it holds for all $\l$ in $[\l]$. Condition (\ref{i})
strengthens the notion that $\Delta$ has `reached equilibrium':
where as the intrinsic metric $q_{ab}$ is `time-independent' on
any non-expanding horizon, on a weakly isolated horizon, the
connection 1-form $\omega$ is also `time-independent'. Since
$\l^a$ is normal to $\Delta$, one can regard $K_a{}^b :=
\nabla_\pback{a} \l^b$ as an analog of the extrinsic curvature of
the null surface $\Delta$. In this sense, on a weakly isolated
horizon, not only the intrinsic metric $q_{ab}$ but also the
extrinsic curvature $K_a{}^b$ is `time independent'; while a
non-expanding horizon approximates a Killing horizon only to
`first order', an isolated horizon approximates it to `first'
\textit{and} `second' order.

We will first make a few remarks to elucidate this Definition and
then work out some of its consequences, including the zeroth law.

\medskip\noindent {(a)} \textit{Remaining rescaling freedom}:
A Killing horizon (with $S^1$-cross-sections) is automatically a
weakly isolated horizon (provided the matter fields satisfy the
energy condition of Definition 1).  Furthermore, given a
non-expanding horizon $\Delta$, one can always find an equivalence
class $[\l ]$ of null-normals such that $(\Delta, [\l ])$ is a
weakly isolated horizon. However, condition (\ref{i}) does not by
itself single out the appropriate equivalence class $[\l]$
uniquely. As indicated in Section \ref{s6.4}, one \textit{can}
further strengthen the boundary conditions and provide a specific
prescription to select the equivalence class $[\l ]$ uniquely.
However, for mechanics of isolated horizons, these extra steps are
unnecessary. In particular, our analysis will not depend on how
the equivalence class $[\l ]$ is chosen. The adverb `weakly' in
Definition 2 emphasizes this point.

\medskip\noindent {(b)} \textit{Surface gravity}: In the case of
Killing horizons $\Delta_{\rm K}$, surface gravity is defined as
the acceleration of the Killing field $\xi$ normal to $\Delta_{\rm
K}$. However, if $\Delta_{\rm K}$ is a Killing horizon for $\xi$,
it is also a Killing horizon for $c\,\xi$ for any positive
constant $c$. Hence, surface gravity is not an intrinsic property
of $\Delta_{\rm K}$, but depends also on the choice of a specific
Killing field $\xi$. (Of course the result that the surface
gravity is constant on $\Delta_{\rm K}$ is insensitive to this
rescaling freedom.) This ambiguity is generally resolved by
selecting a preferred normalization in terms of the structure at
infinity. However, in absence of a \textit{global} Killing field
this strategy does not work and we simply have to accept the
constant rescaling freedom in the definition of surface gravity.
In the context of isolated horizons, then, it is natural to keep
this freedom.

A weakly isolated horizon is similarly equipped with a preferred
family $[\l]$ of null normals, unique up to constant rescalings.
It is natural to interpret $\kappa_{\ls}$ as the surface gravity
associated with $\l$. Under permissible rescalings $\l \mapsto
\tilde\l = c \l$, the surface gravity transforms via:
$\kappa_{(\tilde{\l})} = c \kappa_{\ls}$. Thus, while $\omega$ is
insensitive to the rescaling freedom in $[\l]$, $\kappa_\ls$
captures this freedom fully.  One can, if necessary, select a
specific $\l$ in $[\l]$ by demanding that $\kappa_\ls$ be a
specific function of the horizon parameters which are insensitive
to this freedom, e.g., by setting $\kappa_\ls = (R_\Delta/l^2)$,
where $R_\Delta$ is the horizon radius and $\Lambda = -(1/l^2)$,
the cosmological constant.

\medskip\noindent {(c)} \textit{Zeroth law}: We will now show that
the surface gravity $\kappa_{(\l)}$ is constant on $\Delta$.
Applying the Cartan identity to $(\omega, \l)$ we have:
\begin{equation}\label{eqn:zerothlaw}
  0 = \Lie_{\l}\, \omega = d (\l\cdot\omega) + \l\cdot d\omega.
\end{equation}
However, we have already seen that $\omega$ is curl-free on any
non-expanding horizon. Hence $d (\l\cdot\omega)$ is zero, i.e.,
\begin{equation}\label{eqn:kappaconst}
  \kappa_{(\l)} \= \mbox{const.}
\end{equation}
Thus, weakly isolated horizons have constant surface gravity; the
zeroth law holds on all weakly isolated horizons $(\Delta, [\l])$.
However, as noted above, the precise value of surface gravity
$\kappa_{(\l)}$  depends on the choice of a specific normal $\l$
in $[\l ]$, unless $\kappa_{(\l)}$ vanishes, i.e., $(\Delta, [\l
])$ is an extremal weakly isolated horizon.

\subsection{Symmetries of weakly isolated horizons}
\label{s2.3}

Let us now analyze the symmetries of a weakly isolated horizon.
This analysis will play a key role in the construction of the
horizon angular momentum and energy.

By its definition, a weakly isolated horizon is equipped with
three basic fields: i) the equivalence class $[\l]$ of
null-normals; ii) the intrinsic (degenerate) metric $q_{ab}$ of
signature (0,+), and, iii) the one-form $\omega_a$. Therefore it
is natural to define symmetries of a given weakly isolated horizon
as diffeomorphisms of $\Delta$ which preserve these three fields.

At an infinitesimal level, then, a vector field $\xi^a$ on a
weakly isolated horizon $(\Delta, [\l])$  will be called a
symmetry if
\begin{equation} \Lie_\xi\, \l^a = C \l^a; \quad \quad \Lie_\xi\, q_{ab} \=
0;\quad {\rm and}\quad \Lie_\xi\, \omega \= 0; \end{equation}
for some (possibly vanishing) constant $C$. Now, any vector field
$\xi^a$ on the horizon can be written as a linear combination of
the fields $\l^a$ and $m^a$
\begin{equation}
  \xi^a = A \l^a + B m^a.
\end{equation}
To qualify as a symmetry, the coefficients $A, B$ have to be
constrained appropriately. A simple calculation shows that, if the
surface gravity $\kappa_{(\l)}$ is non-zero, these conditions
reduce to
\begin{equation}
  A  = \mbox{const.}\ \ \ B = \mbox{const.}\, ,
\end{equation}
while if $\kappa_{(\l)}$ is zero, the condition on $A$ is weakened
to
\begin{equation}
  A = C(\phi) +  Ev,
\end{equation}
where $\phi, v$ are given by $\l={\partial}/{\partial v}$,
$m=({1}/R_{\Delta}){\partial}/{\partial \phi}$, and $E$ is a
constant.

Note that, by the definition of weak isolation, $\l^a$ is always
an infinitesimal symmetry of $(\Delta, [\l ])$.  In the generic,
non-extremal case, the only other possible symmetry is the
rotational one. Thus, in this case there are only two
possibilities:\\ i) The symmetry group is two dimensional and
Abelian. In this case metric $q_{ab}$ and the connection 1-form
$\omega_a$ \textit{on the horizon} are stationary, axi-symmetric.
We will refer to these as type I horizons. In this case, we will
be able to introduce a natural notion of angular momentum. The
event horizons of all known stationary black hole solutions are of
type I.\\ ii) The symmetry group is 1-dimensional and corresponds
only to `time' translations along $[\l]$. In this case, at least
one of these fields fails to be axi-symmetric. These are type II
horizons.

In the special, extremal (i.e., $\kl \= 0$) case, the group can be
infinite dimensional.

\subsection{The Maxwell field}
\label{s2.4}

So far, we have focused only on gravitational fields at the
horizon. Let us now allow Maxwell fields and analyze the
implications of the conditions in Definitions 1 and 2.

Recall first that the stress-energy tensor of a Maxwell field
${\bf F}_{ab}$ is given by%
\footnote{The numerical factor $1/2\pi$ ---rather than $1/4\pi$---
is essential to ensure that the first law has the familiar
numerical coefficients even within the family of known solutions.}
\begin{equation}
   T_{ab} = \frac{1}{2\pi} \left[ {\bf F}_{ac} {\bf F}_{b}^{\; c} -
   \frac{1}{4} g_{ab} { \bf F}_{cd} { \bf F}^{cd} \right].
  \end{equation}
Since $R_{ab}\l^a \l^b \= 0$, using field equations at $\Delta$,
we conclude $T_{ab}\l^a \l^b \=0$ which in turn implies ${\bf
F}_{ab}\l^a m^b \= 0$. The condition $R_{ab}\l^a m^b \= 0$ does
not constrain ${\bf F}_{ab}$ any further. Thus, the boundary
conditions imply that the Maxwell field is constrained on $\Delta$
by:
\begin{equation} \label{emAonih} \pback{{\bf F}} \equiv  \pback{d {\bf A}} \=
0 \end{equation}

As in higher dimensions, the electric charge is defined as a
surface integral and conserved because of Maxwell's equations. The
horizon charge $Q_\Delta$ is given by
\begin{equation}  \label{charge} Q_{\Delta} = - \frac{1}{2
\pi}\oint_{S_{\Delta}}\, \dual{\bf F} \end{equation}
and is well-defined because $\dual{\bf F}$, the Hodge-dual of
${\bf F}$, is a 1-form. By contrast, since ${\bf F}$ is a 2-form,
we can not integrate it on a cross-section to obtain a horizon
magnetic charge. (One might imagine integrating ${\bf F}$ over the
whole horizon but this integral vanishes because $\pback{{\bf F}}
\= 0$.)

\textsl{Remark}: Because the first homology of $\Delta$ is
non-trivial, one can define a Aharanov-Bohm charge
$\tilde{Q}_\Delta$:
\begin{equation}\label{bacharge} \tilde{Q}_\Delta =
\frac{1}{2\pi}\oint_{S_\Delta}\, {\bf A} \end{equation}
The integral on the right is `conserved', i.e., is independent of
the cross-section $S_{\Delta}$ on which it is evaluated because
$\pback{\bf F} \= 0$. However, away from $\Delta$, this charge is
not conserved and at infinity it fails to be well-defined because
${\bf A}$ diverges logarithmically.

Finally, let us analyze the electromagnetic scalar potential
$\Phi_{(\l)} := -{\bf A}_a\l^a$.  Since $\omega_a$ is the
gravitational analog of ${\bf A}$, $\Phi_{(\l)}$ can be regarded
as the electromagnetic analog of the surface gravity
$\kappa_{(\l)} \equiv \omega_a \l^a$. Let us first note that since
$\pback{{\bf F}} \= 0$, we can always choose a gauge in which the
vector potential ${\bf A}$ satisfies $\Lie_{\l}\, \pback{{\bf
A}}\= 0$. The standard analysis of Killing horizons strongly
suggests that this is a natural gauge choice on the horizon. A
vector potential ${\bf A}$ satisfying this condition will be said
to be \textit{in a gauge adapted} to $(\Delta, [\l])$. In this
gauge, we have:
\begin{equation} \pback{d}\Phi_{(\l)} \= \Lie_{\l}\, \pback{{\bf A}} \= 0;
\end{equation}
$\Phi_{(\l)}$ is constant on $\Delta$. This is the electromagnetic
counterpart of the zeroth law established above.

\section{Action principle}
\label{s3}

Fix a manifold $\man$, topologically $M\times \Rbar$, with an
inner boundary $\Delta$ which is topologically $S^1\times \Rbar$,
and future and past space-like boundaries $M^\pm$, which are
partial Cauchy surfaces. We will denote the cylinder serving as
the boundary at infinity by $\tau_\infty$. We will assume that the
complement of a compact set of $M$ is diffeomorphic to the
complement of a compact set in $\Rbar^2$; topological
complications, if any are confined to a compact set. We equip the
inner boundary $\Delta$ with an equivalence class of vector fields
$[\l]$ which are transversal to the $S^1$-cross-sections of
$\Delta$ (and where, as before, $\l \sim \l^\prime$ if and only if
they are related by a constant rescaling). Finally, we fix on
$\Delta$ an internal triad $(\l^I, n^I, m^I)$ (with $\l\cdot n \=
-1,\,  m\cdot m \=1$, and all other inner products zero) and raise
and lower its internal indices with a fixed Minkowskian metric
$\eta_{IJ}$ on the internal space.

We will use a first order framework based on (orthonormal)
co-triads
%
$e_I$ and $SO(2,1)$ connections $A_a{}^I$ where $I$ takes values
in the Lie algebra of $SO(2,1)$. These fields will be subject to
certain boundary conditions. On the inner boundary, we will
require: i) $\l^a \= \l^I e_I^a$ belong to $[\l]$ on $\Delta$; ii)
$(\Delta, [\l])$ is a weakly isolated horizon; and, iii) ${\bf A}$
is in an adapted gauge. As mentioned in the Introduction, the
conditions at infinity turn out to be rather subtle because of
peculiarities associated with 2+1 dimensions. As usual, the
conditions should be weak enough so that a large class of
interesting space-times is admissible and strong enough for the
action principle, the phase space and Hamiltonians generating
interesting canonical transformations to be well-defined. In
Appendix B, we present such a choice.

The action for 2+1-dimensional Einstein-Maxwell theory is given
by:
\begin{eqnarray}
    S(e, A, {\bf A}) &=&  \int_{\man}
    \left( e^{I} \wedge F_{I} - \frac{\Lambda}{6}\, \varepsilon^{IJK}
    e_{I} \wedge e_{J} \wedge e_{K} \right) - \frac{1}{2}
    \int_{\tau_\infty}\, e^{I} \wedge A_{I}
    \nonumber \\
    &-& \frac{1}{4 \pi} \int_{\man} {\bf F} \wedge
    \dual{\bf F}+ \frac{1}{4\pi} \int_{\tau_\infty}
    \dual{\bf F} \wedge {\bf A} \label{action}
\end{eqnarray}
Here, $F_{I}$ is the curvature of the gravitational connection
$A^I$, ${\bf F}$ the curvature of the electromagnetic connection
${\bf A}$ and $\dual{\bf F}$ its Hodge dual. All integrals should
be understood as suitable limits of integrals evaluated on finite
regions of $\man$ and their boundaries as the regions expand to
fill $M$ and boundaries tend to $\tau_\infty = S_\infty\times
\Rbar$. Then, with our boundary conditions the action is finite
and its variations are well-defined on the entire space of
histories under consideration. In contrast to the asymptotically
flat situation (in 3+1 dimensions) considered in earlier papers
\cite{afk,abl2}, here the surface terms at infinity are essential
to ensure that the action is finite.

Let us vary the action keeping fields fixed on the initial and
final surfaces $M_\pm$. Since the calculation is closely analogous
to that in 3+1 dimensions \cite{abf}, we will only sketch the main
steps. We have:
  \begin{equation}
   \delta S_{\rm grav} = {\rm bulk \; \; terms}
   + \int_{\Delta} e^{I} \wedge \delta A_{I},
  \end{equation}
where the bulk terms just provide the equations of motion,
provided the surface terms vanish. There is no surface term at
infinity because of the asymptotic conditions of Appendix B. Let
us examine the surface term at the horizon. It can be further
simplified: our boundary conditions imply that the pull-back
$\pback{A^I}$ to the horizon of the gravitational connection $A^I$
is necessarily of the form
\begin{equation}  \label{Aonih} \pback{A}^I \=  \omega m^I + C \l^I
\end{equation}
where $C$ is a 1-form on $\Delta$ which is annihilated by $\l^a$.
(For a Newman-Penrose type derivation, see Appendix A.4.) Hence,
\begin{equation}
\delta S_{\rm grav} = {\rm bulk \; \; terms} + \int_{\Delta} e^{I}
\wedge (\delta \omega) m_{I} ,
\end{equation}
where we used the fact that the \textit{internal} triad is kept
fixed on $\Delta$. Now, since $\delta\l^a \= c_\delta \l^a$ for
some constant $c_\delta$,  $ {\cal L}_\l \omega \= 0$ in each
history, and the variation $\delta \omega$ vanishes on the initial
and final cross-sections of $\Delta$ (i.e., on the intersections
of $\Delta$ with $M_\pm$), we conclude $\delta\omega \= 0$ on all
of $\Delta$. Thus, all the gravitational surface terms vanish
under permissible variations.

The situation with the electromagnetic terms is analogous. We
have:
\begin{equation}
\delta S_{\rm Maxwell} = {\rm bulk \; \; terms} -
\frac{1}{2\pi}\int_{\Delta} \delta {\bf A}\wedge \dual{\bf F}
\end{equation}
Since ${\bf A}$ is assumed to be in an adapted gauge, ${\cal L}_\l
\pback{{\bf A}} \= 0$. Again, since $\delta\l^a \= c_\delta \l^a$
for some constant $c_\delta$ and the variation $\delta \pback{\bf
A}$ vanishes on the initial and final cross-sections of $\Delta$,
the surface term vanishes.

Thus the variations of the action $S(e,A, {\bf A})$ are
well-defined and just yield the Einstein-Maxwell equations.

\section{Covariant phase space and the first law}
\label{s4}

In this section we will construct the covariant phase space,%
\footnote{These derivations were first carried out using the
Legendre transform and the resulting canonical phase space.
However, in that framework, a few conceptual complications arise
in the intermediate steps which are finally irrelevant for our
results, and, furthermore, calculations are significantly more
complicated. Therefore we decided to use the covariant phase space
in this presentation.}
use it to introduce the notion of angular momentum and energy on
$\Delta$, and obtain the first law. This approach was used in
higher dimensional discussions of black hole mechanics
\cite{afk,abl2}, and, as in that discussion, our first law will
arise as a consistency condition for the time-evolution to be
Hamiltonian. Therefore we will only sketch the main steps and
emphasize the peculiarities of 2+1 dimensions.

\subsection{Phase space}
\label{s4.1}

To be able to define angular momentum, we will now restrict
ourselves to type I isolated horizons of section \ref{s2.3}. Thus,
in addition to the structures introduced in the beginning of
Section \ref{s3}, we now equip the inner boundary $\Delta$ with a
vector field $\varphi^a$ such that its affine parameter $\varphi$
runs from $0$ to $2\pi$. Since $\Delta$ is assumed to be of type
I, the intrinsic metric $q_{ab}$ and the 1-form $\omega$ are
Lie-dragged by $\varphi^a$. (Note that this condition is imposed
\textit{only} at $\Delta$;  we do \textit{not} ask that there be
an axial Killing field outside, even in a neighborhood of
$\Delta$.) For simplicity, we will also assume that $\varphi^a$ is
tangential to the intersections $S_\Delta^\pm$ of $\Delta$ with
the past and future surfaces $M^\pm$. Finally, it is convenient to
introduce two scalar fields $\psi$ and $\chi$ on $\Delta$ which
serve as `potentials' for the surface gravity $\kappa_{(\l)}$ and
its electro-magnetic analog $\Phi_{(\l)}$ via: i) ${\cal L}_{\l}\,
\psi \= \kappa_{(\l)}$ and ${\cal L}_{\l}\, \chi \= \Phi_{(\l)}$;
and, ii) $\psi$ and $\chi$ vanish on $S_\Delta^{-}$.

\begin{figure} \label{f1}
  \begin{center}
  \includegraphics[height=5cm]{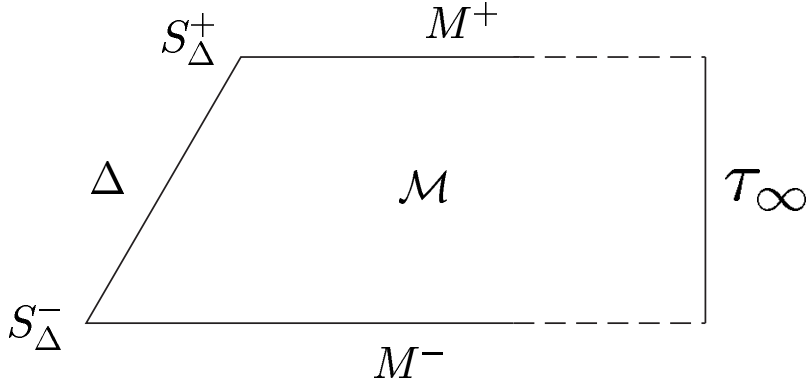}
    \bigskip
  \caption{The region of space-time $\man$ under consideration has an
  internal boundary $\Delta$ and is bounded by two partial Cauchy
  surfaces $M^{\pm}$ which intersect $\Delta$ in the $2$-spheres
  $S^{\pm}_\Delta$ and extend to spatial infinity where they cross
  the cylinder $\tau_{\infty}$. }
  \end{center}
\end{figure}

Our covariant phase space $\Gamma$ will consist of
\textit{solutions} $(A_{I},e^{I},{\bf A})$ to the Einstein-Maxwell
equations, satisfying the above boundary conditions. As usual, to
construct the symplectic structure on $\Gamma$, we begin with the
(anti-symmetrized) second variation of the action (\ref{action}).
Applying the equations of motion to this second variation, one
finds that the integral over $\man$ reduces to surface terms at
$M^{\pm}$ and at $\Delta$. The surface term at $\tau_{\infty}$
vanishes because of the asymptotic fall-off conditions.
Furthermore, expressed in terms of $\psi$ and $\chi$, the surface
term at $\Delta$ turns out to be exact and thus reduces to a pair
of integrals on $S_{\Delta}^{\pm}$. The integral over $M^{+}$,
together with its surface term at $S_{\Delta}^{+}$ is then taken
to define the symplectic structure $\Omega$  on $\Gamma$:
\begin{eqnarray}
   \Omega{\mid_{(A,e,{\bf A})}}\,(\delta_{1},\delta_{2}) &=&  -\int_{M}
   \left( \delta_{1} A^{I} \wedge \delta_{2} e_{I} - \delta_{2} A^{I} \wedge
   \delta_{1} e_{I} \right) - \nonumber\\
   & - & \oint_{S_\Delta} \left(\delta_1
   \psi\, \delta_2 m - \delta_2 \psi\, \delta_2 m \right) + \\
   &+& \frac{1}{2 \pi} \int_{M} \left(\delta_{1}
   {\bf A} \wedge \delta_{2} \dual{\bf F} - \delta_{2}
   {\bf A} \wedge \delta_{1} \dual{\bf F} \right) + \nonumber \\
   & + & \frac{1}{2\pi} \oint_{S_\Delta}
   \left(\delta_1 \chi \, \delta_2 \dual{\bf F} - \delta_2 \psi \, \delta_1
    \dual{\bf F} \right)\nonumber
   \label{sform}
  \end{eqnarray}
for any two tangent vectors $\delta_{1}$ and $\delta_{2}$ at the
phase space point $(A,e, {\bf A})$.

One can verify that, with our boundary conditions, the integral
converges in spite of the logarithmic divergences at infinity and,
because of field equations, it is conserved in spite of the
presence of internal boundaries. More precisely, given a general
solution to the field equations and for general solutions
$\delta_1$ and $\delta_2$ to the linearized equations on $\man$,
the integral (\ref{sform}) evaluated on a partial Cauchy slice $M$
is well-defined and independent of the choice of that slice
\cite{afk,abl2}. Note that this conservation would \textit{not}
hold had we left out the boundary term. The integral of the
`symplectic current' constructed from the bulk terms across
$\Delta$ is compensated by the difference between the boundary
terms evaluated at $M_\pm$ \cite{afk}.

\textsl{Remark}: As in higher dimensions, the term `symplectic
structure' is somewhat of a misnomer because in the covariant
phase space framework, $\Omega$ can have degenerate directions
which correspond to infinitesimal gauge transformations. However,
because the first homology of $\man$ is non-trivial, there is now
an interesting subtlety involving the electromagnetic potential.
Denote by $\tilde\phi_a$ the closed 1-form on $\man$ which can be
\textit{locally} expressed as the exact differential $\partial_a
\phi$. Then, $\delta {\bf A}_a = \tilde\phi_a$ is a pure gauge
connection from a \textit{space-time} perspective because $\delta
{\bf F} = d \tilde\phi =0$. However, this $\delta{\bf A}$ does
\textit{not} belong to the kernel of the symplectic structure
because it is not exact. Therefore, from a \textit{phase space}
perspective, this is not a pure gauge connection. From now on, we
will use the phase space notion of gauge. Thus, an expression will
be said to be gauge invariant if it is invariant under ${\bf A}
\mapsto {\bf A} + df$ for a smooth function $f$.

\subsection{Angular momentum}
\label{s4.2}

Let $\phi^a$ be a vector field on $\man$ with closed orbits whose
affine parameter runs between $0$ and $2\pi$ such that it is a
rotational Killing vector of the asymptotic metric at infinity and
coincides with the fixed rotational symmetry vector field
$\varphi$ on $\Delta$. ($\phi^a$ is \textit{not} required to be a
Killing field in the bulk; indeed general metrics in our phase
space do not admit \textit{any} Killing field.) Diffeomorphisms
generated by $\phi^a$ naturally induce a vector field
\begin{equation} \delta_{\phi} = ({\cal L}_\phi\, A,\, {\cal L}_\phi\, e, \,
{\cal L}_\phi\, {\bf A}) \end{equation}
on the phase space $\Gamma$. It is natural to ask whether it
preserves the symplectic structure. As on any phase space, the
answer is in the affirmative if and only if the 1-form
$X^{(\phi)}$ on $\Gamma$ defined by
\begin{equation} \label{X} X^{(\phi)} (\delta)  = \Omega (\delta,
\delta_\phi) \end{equation}
is exact, where $\delta$ is an arbitrary vector field on $\Gamma$.
A direct calculation of the right hand side of (\ref{X}) shows
that this is indeed the case: $X^{(\phi)} = d J^{(\phi)}$ where
the phase space function $J^{(\phi)}$ is given, up to an additive
constant, by:
\begin{eqnarray}\label{J} J^{(\phi)} &=& \oint_{S_{\Delta}} \left[(\varphi
\cdot \omega) m - \frac{1}{2 \pi} (\varphi \cdot {\bf A})
\dual{\bf F}\right] + \nonumber\\
& + &  \oint_{S_\infty}\, \left[(\varphi \cdot e_I) A^I + -
\frac{1}{2 \pi} (\varphi \cdot {\bf A}) \dual{\bf F}\right] \\
&=& J_\Delta - J_\infty\nonumber
\end{eqnarray}
(Because of the absence of a background geometry, the Hamiltonians
generating space-time diffeomorphisms in the \textit{covariant}
phase space consist only of boundary terms.) The requirement that
$J^{(\phi)}$ must vanish in the non-rotating BTZ solution implies
that the undetermined constant must be zero. Finally, in spite of
the fact that the electromagnetic potential ${\bf A}$ appears
explicitly, the expression is gauge invariant.

The integral $J_\infty$ at infinity is the total angular momentum
of the system, including contributions from matter fields outside
$\Delta$. Note that, in contrast to the situation in 3+1
dimensions, this surface term contains a contribution from the
Maxwell field. The evaluation of the surface term is delicate. As
before, one has to first evaluate the integral on the exterior
boundary of a finite region of a partial Cauchy surface and then
take the limit to infinity. In the limit, the contributions from
the Maxwell and the gravitational parts diverge individually but
the sum is finite.%
\footnote{It is because of such subtleties that the expression for
the total mass and angular momentum in the general charged,
rotating case had been unavailable in 2+1 gravity \cite{sc}.}
Finally, it is natural to interpret the horizon integral
$J_\Delta$ as the horizon angular momentum. As in higher
dimensions \cite{abl2}, this interpretation is supported by
various properties. In particular, if the space-time admits a
rotational Killing field $\phi^a$ in a neighborhood of $\Delta$,
then
\begin{equation} J^{\rm grav}_\Delta \equiv J^{\rm Komar}_\Delta = -
\frac{1}{2}\, \oint_{S_\Delta}\, \dual{d\phi}\, ,\end{equation}
where $J^{\rm grav}_\Delta$ is the gravitational part of the
horizon angular momentum $J_\Delta$ in (\ref{J}). It is
straightforward to check that, in the BTZ solutions, $J_\Delta =
J$, where $J$ is the parameter in the BTZ metric. Also, as one
might expect from the presence of a \textit{global} Killing field
$\phi$, in these solution $J_\Delta = J_\infty$ so that the
Hamiltonian $J^{(\phi)}$ generating the diffeomorphism along
$\phi$ vanishes identically. From general symplectic geometry
considerations, it follows that this result holds also on the
entire connected component of \textit{axi-symmetric} solutions
containing the BTZ  solution. In the general non axi-symmetric
case, on the other hand $J^{(\phi)}$ is non-zero and represents
the angular momentum in the Maxwell field outside the horizon.
Finally, the electromagnetic part of the horizon term $J_\Delta$
can be expressed in terms of the electric charge $Q$ and the
Aharanov-Bohm charge $\tilde{Q}$ of (\ref{bacharge}) :
\begin{equation} J_\Delta^{\rm em} = - \frac{1}{2\pi} \oint_{S_\Delta}\,
(\varphi \cdot \dual{\bf F})\, {\bf A} = - \frac{1}{2\pi} Q
\tilde{Q} \end{equation}

\subsection{Energy and the first law}
\label{s4.3}

Following the strategy adopted for defining angular momentum, it
is natural to define horizon energy as the appropriate surface
term in the expression of the Hamiltonian generating
time-evolution. Let us therefore begin by introducing, in each
history, an `evolution vector field' $t^a$. To qualify as a `time
translation', $t^a$ will be required to be a generator of an
appropriate symmetry at the two boundaries: we will assume that it
approaches a \textit{fixed} time translation at spatial infinity
and has the form $t^a = c_{\t}\l^a - \Omega_{\t}\varphi^a$ on
$\Delta$, where $c_{\t}$ and $\Omega_{\t}$ are constants on
$\Delta$. It turns out to be necessary to allow $t^a$ to vary from
one space-time to another; in the numerical relativity
terminology, the vector field is allowed to be `live'. The
asymptotic value of $t^a$ at spatial infinity is independent of
the choice of history and defines a fixed time translation Killing
field of the asymptotic metric. On the horizon, on the other hand,
$c_\t$ and $\Omega_\t$ are allowed to vary from one space-time to
another. For example, for physical reasons, in the non-rotating
BTZ solution, we would like $\Omega_\t$ to vanish, while in the
rotating case we would like it to be non-zero. As we will see,
this generalization is essential to obtain a well-defined
Hamiltonian as well as the first law.

The evolution field $t^a$ induces a vector field $\delta_{t}$ on
the phase space, given by
\begin{equation} \delta_{t} = ({\cal L}_t \, A,\, {\cal L}_t \, e, \, {\cal
L}_t\, {\bf A}), \end{equation}
and representing infinitesimal time evolution. The key question
now is whether this evolution is Hamiltonian, i.e., if $\delta_t$
preserves the symplectic structure $\Omega$ on $\Gamma$. As usual,
this is the case if and only if the one form $X^{(t)}$ on $\Gamma$
defined by
\begin{equation}\label{Xt1}
  X^{(t)}(\delta) = \Omega \left( \delta , \delta_{t} \right)
\end{equation}
is closed.

We can evaluate the right-hand side of (\ref{Xt1}) using the
expression (\ref{sform}) of the symplectic structure and simplify
it using equations of motion, conditions (\ref{Aonih}) and
(\ref{emAonih}) on the gravitational and electromagnetic
potentials, and the fact that, on the horizon, $\delta \l \hat{=}
c_{\delta}\l$ for some constant $c_\delta$. The resulting
expression again involves only integrals at the boundary of
space-time $\man$.
\begin{eqnarray}\label{Xt2}
   X^{(t)}(\delta) = X^{(t)}_{\infty}(\delta)
   - \kappa_{(t)} \delta a_{\Delta} -
   \Omega_{(t)} \delta J_{\Delta} - \Phi_{(t)} \delta
   Q_{\Delta},
  \end{eqnarray}
where $X^{(t)}_\infty (\delta)$ involves only fields at infinity;
$\kappa_{(t)}$ and $\Phi_{(t)}$ are, respectively, the surface
gravity and electric potential on $\Delta$, both associated with
$c_{(t)}\l$. Using boundary conditions at infinity, we can express
the term $X^{(t)}_{\infty}(\delta)$ as an exact variation. As in
the case of angular momentum, the actual evaluation of this
surface term is somewhat delicate: the gravitational and the
electromagnetic terms diverge individually; it is only the sum
that is finite. The final result is:
\ba
  X^{(t)}_{\infty}(\delta) & = & \delta E^t_\infty, \quad {\rm
  with} \quad \nonumber \\
  E^t_\infty & = &\frac{1}{4} \Lambda Q^{2}
  \omega^{2} - \half Q^{2} \ln
  \frac{\bar{r}_{0}}{\bar{l}}(1-\Lambda \omega^{2}),
  \label{Einf}
\ea
where the parameters $Q$, $\omega$ and $\bar{r}_{0}$ can be read
off from the asymptotic behavior specified in Appendix B.
$E^t_\infty$ is the \textit{energy at infinity} corresponding to
the asymptotic time translation $t^a$. (Again, the freedom to add
a constant is eliminated by requiring that $E^t_\infty$ should
yield $M$ when $t^a$ is chosen to be the standard time translation
in non-rotating BTZ space-times.)

{}From (\ref{Xt2}) and (\ref{Einf}) we conclude that the evolution
along $t^a$ is Hamiltonian if and only if the \textit{horizon
term} in (\ref{Xt2}) is an exact variation, i.e., if and only if
there exists a function $E_{\Delta}^{t}$ on the phase space,
constructed from fields at the horizon, such that
  \begin{equation}
   \delta E_{\Delta}^{t} = \kappa_{(t)} \delta a_{\Delta} +
   \Omega_{(t)} \delta J_{\Delta} + \Phi_{(t)} \delta
   Q_{\Delta}. \label{1law}
  \end{equation}
It is natural to identify $E_{\Delta}^{t}$ as the \textit{horizon
energy}. Remarkably, (\ref{1law}) is precisely the statement of
the first law. Thus, \textit{the first law (\ref{1law}) is the
necessary and sufficient condition that the time evolution
generated by the live vector field $t^a$ on $\man$ is
Hamiltonian}.

Not every live vector field $t^a$ considered above satisfies this
condition. A vector field which does will be said to be
\textit{admissible}. We will show in the next section that there
exists an \textit{infinite} number of admissible vector fields,
whence there is an infinite family of first laws. A natural
question is whether one can make a canonical choice, using our
knowledge of known exact solutions. We will show that the answer
is in the affirmative. The horizon energy defined by this
canonical live vector field will be called the \textit{horizon
mass}.

\textsl{Remark}: In contrast to the asymptotically flat case
treated in higher dimensions \cite{afk,abl2}, in stationary
space-times, the expression (\ref{Einf}) of the energy at infinity
does not agree with the Komar integral. In fact, the Komar
integral now diverges, while our expression is finite.

\section{Horizon mass}
\label{s5}

In this section, we will first introduce a systematic procedure to
construct admissible vector fields and then use our knowledge of
stationary, axi-symmetric black hole solutions to introduce
preferred admissible vector fields on \textit{all} space-times in
the phase space $\Gamma$.

\subsection{Admissible vector fields}
\label{s5.1}

Note first that (\ref{1law}) implies that $t^a$ is an admissible
vector field only if $E_{\Delta}^{t}, \kappa_{(t)}$,
$\Omega_{(t)}$ and $\Phi_{(t)}$ are all functions \textit{only} of
the horizon parameters $(a_{\Delta}, J_{\Delta}, Q_{\Delta})$.
Furthermore (\ref{1law}) implies that the following rather
stringent condition must be met at the horizon:
\begin{equation} \label{cond1}
  \frac{\partial \kappa_{(t)}}{\partial J_{\Delta}} =
  \frac{\partial \Omega_{(t)}}{\partial a_{\Delta}}.
\end{equation}
We will turn the argument around and use this equation to
construct admissible vector fields. Let us begin by fixing a
`suitably regular' function $\kappa_{0}(a_{\Delta}, J_{\Delta},
Q_{\Delta})$ of the horizon parameters. Now, given a general
solution, the surface gravity $\kl$ of the null generator $\l$
will not equal $\kappa_{0}$. However, there will be a unique
constant $c$ such that $\kappa_{(c\l)} = \kappa_0$. Next, we find
a constant $\Omega_{(t)}$ by integrating (\ref{cond1}) with
respect to $a_\Delta$:
\begin{equation}
  \Omega_t = \int_{\infty}^{a} \frac{\partial \kappa_{0}}{\partial
  J_{\Delta}}\, {d}a_{\Delta} \,+\, F(Q_{\Delta},J_{\Delta}),
\end{equation}
where $F$ is an arbitrary function of the two parameters. (The
qualification `suitably regular' above is meant to ensure that the
integral on the right is well-defined.) Finally, we can fix the
arbitrariness in $\Omega_{(t)}$  by imposing the following
physical requirement:
$$\lim_{\stackrel{J_{\Delta}, Q_{\Delta} =
\mbox{{\scriptsize}{const.}}}{a_{\Delta} \rightarrow \infty}}
\Omega_{(t)} = 0.$$
Now, in any given solution in the phase space, we choose any
evolution vector field $t^a$ such that it tends to the fixed
asymptotic time-translation at infinity and satisfies $t^a \= c
\l^a - \Omega_{(t)} \varphi^a$ on $\Delta$. It is straightforward
to check that, by construction, this evolution vector field is
admissible if the Maxwell field of the solution under
consideration vanishes on $\Delta$.

If the Maxwell field on $\Delta$ is non-zero, we must also ensure
that the Maxwell gauge is fixed appropriately for (\ref{1law}) to
hold. Recall first that in an adapted gauge, the Maxwell potential
${\bf A}$ is such that $\Phi_\l \= -{\bf A}_a \l^a$ is constant on
$\Delta$. However, the \textit{value} of the constant, i.e., its
possible dependence on the horizon parameters, is still completely
unconstrained. Equation (\ref{1law}) imposes severe restrictions
on this choice: $\Phi_{(t)} \equiv \Phi_{(c\l)}$ must satisfy
\begin{eqnarray} \label{cond2}
  \frac{\partial \Phi_{(t)}}{\partial a_{\Delta}} =
  \frac{\partial \kappa_{(t)}}{\partial Q_{\Delta}},\nonumber \\
  \frac{\partial \Phi_{(t)}}{\partial J_{\Delta}} =
  \frac{\partial \Omega_{(t)}}{\partial Q_{\Delta}}.
\end{eqnarray}
Again, we can just \textit{use} this condition to constrain
$\Phi_{(t)}$: setting $\kappa_{(t)} = \kappa_0$ and using
$\Omega_{(t)}$ determined above, we can simply integrate these
equations to determine $\Phi_{(t)}$ up to an additive function
$F(Q)$ of the charge, $Q_{\Delta}$. In 3+1 dimensions, there was a
natural way to fix this freedom \cite{abf,abl2}: One could just
impose the physical requirement that $\Phi_{(t)}$ should vanish in
the limit of large areas, with fixed charge and angular momentum.
Unfortunately, in 2+1 dimensions this strategy is \textit{not}
viable because now, in presence of a non-zero charge, the
potential \textit{diverges} at spatial infinity! Therefore now
$\Phi_{(t)}$ is not completely determined on $\Delta$. The only
physical restriction we impose on $F(Q)$ is through
\begin{equation}
\lim_{\stackrel{J_{\Delta}, a_{\Delta} =
\mbox{{\scriptsize const.}}}{Q_{\Delta} \rightarrow 0}}
\Phi_{(t)} = 0,\label{PhiQ}
\end{equation}
which only determines the value of $F(Q)$ at $Q= 0$.

Note, however, that the remaining freedom is irrelevant for the
purpose of defining admissible vector fields: the vector fields
$t^a$ constructed above are admissible for \textit{every} choice
of $\Phi_{(t)}$ satisfying (\ref{cond2}). However, the choice of
$\Phi_{(t)}$ will, in general, enter the expression of the horizon
energy $E^t_\Delta$ which is obtained by integrating (\ref{1law}).

\subsection{Preferred admissible vector fields}
\label{s5.2}

In this section, we will indicate how one can use the known
solutions to fix $\kappa_0$ and $\Phi_{(t)}$ in a `canonical
fashion'. The resulting $E^t_\Delta$ can be naturally interpreted
as the horizon mass. Several subtleties arise in presence of a
non-zero charge and angular momentum. Therefore we will divide the
discussion into three cases.

\subsubsection{The case with ${\bf F}_{ab}\= 0$}
\label{s5.2.1}

Let us suppose that the Maxwell field vanishes on the horizon.
Then we only have to choose a function $\kappa_0$ of the horizon
parameters $a_\Delta, J_\Delta$. However, in this case, there is a
\textit{unique} BTZ black hole solution for each choice of these
two parameters. Therefore, it is natural to set $\kappa_0 =
\kappa_{(t)}^{\rm BTZ}$, where $t^a$ is the canonical
time-translation Killing field of the BTZ black hole:
  \begin{equation}
\kappa_0 \equiv  \kappa_{(t)} = - \frac{\Lambda a_{\Delta}}{2 \pi} - \frac{2
   \pi J_{\Delta}^{2}}{a_{\Delta}^{3}}.
  \end{equation}
Our construction of Section \ref{s5.1} implies
  \begin{equation}
   \Omega_{(t)} = \frac{2 \pi J_{\Delta}}{a_{\Delta}^{2}}.
  \end{equation}
We can now integrate out the first law to obtain the expression of
the horizon energy up to an undetermined additive constant. We
eliminate this freedom through the physical requirement:
$\lim_{a_{\Delta} \rightarrow 0} E_{\Delta}^{t} = 0$ for
non-rotating isolated horizons. The resulting horizon mass is
given as a function of the horizon parameters as:
\begin{equation} \label{mass1}
   M_\Delta = - \frac{1}{4 \pi} \Lambda a_{\Delta}^{2} +
   \frac{\pi J_{\Delta}^{2}}{a_{\Delta}^{2}}.
\end{equation}
The functional form of $M_\Delta$  is the same as that in the BTZ
family. However, (\ref{mass1}) was not simply postulated but
derived systematically from Hamiltonian considerations and applies
to \textit{all} isolated horizons including those which may admit
electromagnetic radiation in the exterior region, away from
$\Delta$. In presence of such radiation, $M_\Delta$ will not equal
the mass at infinity.

\subsubsection{Charged, non-rotating horizons}
\label{s5.2.2}

Next, let us consider non-rotating horizons with electric charge.
For brevity, we will treat this as a sub-case (corresponding to
$\omega = 0$) of the Cl\'{e}ment solution. The metric and the
Maxwell field of this solution were given in section \ref{s2.1}.
The corresponding electromagnetic potential, satisfying the
boundary conditions of Appendix B, is given by
\begin{equation}
  {\bf A} = Q \ln(\frac{r}{l})dv .
\end{equation}
Using the Killing field $t =\partial_{v}$ as the evolution field,
it is again natural to set $\kappa_0 = \kappa_{(t)}$ and
$\Phi_{(t)} \= -{\bf A}_a t^a$. Thus, we now have:
  \begin{equation}\label{1}
\kappa_0 \equiv  \kappa_{(t)} = - \Lambda \frac{a_{\Delta}}{2 \pi} -
   \frac{Q_{\Delta}^{2}}{2a_{\Delta}}.
  \end{equation}
and
  \begin{equation} \label{2}
   \Phi_{(t)} = - Q_{\Delta} \ln \frac{a_{\Delta}}{a_{0}},
  \end{equation}
where $a_{0}=2\pi l$. Thus, we have used our boundary conditions
at spatial infinity to determine $\Phi_t$ uniquely in these
solutions. Since angular momentum vanishes, $\Omega_t =0$

For general non-rotating weakly isolated horizons, therefore, we
select the canonical evolution vector fields and electromagnetic
scalar potential by demanding that the dependence of $\kappa_0$
and $\Phi_t$ on $a_\Delta$ and $Q_\Delta$ be fixed as in Equations
(\ref{1}) and (\ref{2}), and $\Omega_t$ should vanish. Then, it is
straightforward to integrate the first law to obtain the horizon
mass. We obtain:
  \begin{equation}
   M_{\Delta} = - \frac{\Lambda a_{\Delta}^{2}}{4 \pi} -
   \frac{1}{2} Q_{\Delta}^{2} \ln \frac{a_{\Delta}}{a_{0}}.
  \end{equation}
Again, this formula now holds for arbitrary non-rotating weakly
isolated horizons $\Delta$.

\subsubsection{Charged rotating black hole.}
\label{s5.2.3}

Finally, let us consider the general case. We can now use the
general Cl\'{e}ment solution to fix $\kappa_0$ and $\Omega_{(t)}$.

As in the charged, non-rotating case, we need to specify the
electromagnetic vector potential ${\bf A}_a$. The potential
satisfying our boundary conditions as well as conditions
(\ref{cond2}) which are necessary for the first law to hold is
given by%
\footnote{The  electromagnetic potential used by Cl\'{e}ment
consists only of the first term. This does satisfy our boundary
conditions at spatial infinity and is also in an adapted gauge on
$\Delta$. However, the resulting $\Phi_{(t)}$ does \textit{not}
satisfy conditions (\ref{cond2}). Therefore, we have made a
suitable gauge transformation by adding the second term.}:
   \begin{equation}
    {\bf A} = Q \ln \frac{r}{\bar{l}} (dv - \omega
    d\phi) + \frac{1}{2} Q \omega^{2} \Lambda dv.
   \end{equation}
Next, let us express the horizon parameters $a_\Delta, Q_\Delta$
and $J_\Delta$ in terms of the parameters $\omega, Q, \bar{r}_0$
that appear in the solution:
\ba
a_{\Delta} &=& \frac{2 \pi r_{\Delta}}{\sqrt{1+\omega^{2}\Lambda}}
\nonumber\\
Q_\Delta &=& Q\nonumber\\
J_{\Delta} &=& - \omega \left( \frac{\Lambda a_{\Delta}^{2}}{2
    \pi} + \frac{1}{2}Q_{\Delta}^{2} + Q_{\Delta}^{2} \ln
    \frac{a_{\Delta}}{2 \pi l} \right),
\ea
where $r_{\Delta}$ is given by $N(r=r_{\Delta})=0$. (The parameter
$\bar{r}_0$ enters the expression of the area through this
condition.)

Now we can calculate the surface gravity $\kappa_{(t)}$ and the
electric potential $\Phi_{(t)}$ corresponding to the stationary
Killing field $t^a$ of the Cl\'{e}ment solution:
   \begin{eqnarray}
    \kappa_{(t)} & = & -\left( \frac{\Lambda a_{\Delta}}{2\pi} +
    \frac{Q_{\Delta}^{2}}{2a_{\Delta}} \right)
    (1+\omega^{2}\Lambda), \\
    \Phi_{(t)} & = & - Q_{\Delta} \ln \frac{a_{\Delta}}{a_{0}}
    (1+\omega^{2}\Lambda) - \frac{1}{2} Q_{\Delta} \omega^{2} \Lambda,
   \end{eqnarray}
where, as before, $a_{0}=2\pi l$.

In a general solution in the phase space, then, we set $\kappa_0 =
\kappa_{(t)}$ given above. Our procedure of Section \ref{s5.1}
provides the required $\Omega_{(t)}$:
\begin{equation} \Omega_{(t)}  =  -N^{\phi}(r=r_{\Delta}) = -\omega
    \Lambda . \end{equation}
The triplet $\kappa_{(t)}, \Omega_{(t)}, \Phi_{(t)}$ can now be
used to construct preferred admissible vector fields and by
integrating the corresponding first law, we obtain the expression
of the horizon mass:
   \begin{equation}\label{mass3}
    M_{\Delta} = -\frac{\Lambda a_{\Delta}^{2}}{4\pi}
    -\frac{1}{2} Q_{\Delta}^{2} \ln \frac{a_{\Delta}}{a_{0}} + \frac{1}{2}
    \frac{\Lambda J_{\Delta}^{2}}{\frac{\Lambda
    a_{\Delta}^{2}}{2\pi} + \frac{1}{2}Q_{\Delta}^{2} + Q_{\Delta}^{2}
    \ln \frac{a_{\Delta}}{a_{0}}}
   \end{equation}
where we have again eliminated an undetermined constant by
requiring that every non-rotating, uncharged horizon should have
vanishing mass in the limit of vanishing area.  This is our
\textit{general} expression of the horizon mass.

Finally, we can compare our formula for the energy of the horizon
with the energy at infinity ---Eq.(\ref{Einf}). It is easy to
check that in the case of Cl\'ement's solution, the two
expressions are equal to each other, just as one might expect from
results in 3+1 dimensions. General symplectic arguments
\cite{abl2,abk} now imply that this equality between our horizon
mass and the mass at infinity must continue to hold for
\textit{all} stationary space-times in the connected component of
the phase-space containing Cl\'ement (or, equivalently, BTZ)
solutions.

%

\bigskip

\section{Horizon geometry}
\label{s6}

In this section, we examine geometrical structures on $\Delta$ and
analyze their interplay with the field equations. As mentioned in
the Introduction, this section can be read independently of the
last three.

The section is divided into four parts. In the first, we will show
that every non expanding horizon is naturally equipped with an
intrinsic derivative operator $\D$. In the second, we will turn to
weakly isolated horizons and, using field equations, isolate the
freely specifiable data on $\Delta$. In the third, we will show
that every \textit{non-extremal} weakly isolated horizon admits a
natural foliation (irrespective of whether it is axi-symmetric,
i.e., type I in the terminology of section \ref{s2.3}). In the
last sub-section we strengthen the definition of weak isolation to
introduce the notion of isolated horizons. While a non-expanding
horizon $\Delta$ can be made weakly isolated by suitably choosing
$[\l]$ in infinitely many inequivalent ways, generically, it
admits a unique $[\l]$ which makes it isolated.

\subsection{A natural derivative operator}
\label{s6.1}

Let $\Delta$ be a non-expanding horizon. Had it been  space-like
or time-like,  its intrinsic metric would have selected a unique
(torsion-free) derivative operator. However, since it is null,
there are infinitely many derivative operators which are
compatible with it. Nonetheless, because $\Delta$ is expansion and
shear-free, as in higher dimensions \cite{afk}, the full
space-time derivative operator $\nabla$ induces a preferred
intrinsic derivative $\D$ on it. Given a vector field $X^a$ or a
1-form $f_a$, on $\Delta$, we have:
\begin{equation}\label{D} Y^a \D_a X^b \= Y^a \nabla_a {\bar{X}}^b, \quad
{\rm and} \quad Y^a Z^b \D_a f_b \= Y^a Z^b \nabla_a {\bar{f}}_b
\, ,\end{equation}
where $Y^a, Z^a$ are arbitrary vector fields tangential to
$\Delta$ and $\bar{X}^a$ and ${\bar{f}}_a$ are arbitrary smooth
extensions of $X^a$ and $f_a$ to a space-time neighborhood of
$\Delta$. It is easy to check that $\D$ is well-defined: the right
hand sides of the two equations are independent of the choice of
extension and the right hand side of the first equation is again
tangential to $\Delta$. Since $\nabla_a g_{bc} =0$ in space-time,
$\D_a q_{bc} \=0$ on $\Delta$; as expected, $\D$ is compatible
with $q_{bc}$.

What information does $\D$ have beyond that contained in the
degenerate metric $q_{ab}$ on $\Delta$? The action of $\D$ on
tensors is completely determined by that on all 1-forms defined
intrinsically on $\Delta$. Let $f$ be a 1-form satisfying $f\cdot
\l \= 0$ and ${\cal L}_\l f \=0$. Then it is easy to verify that
the action of $\D$ on $f$ can be expressed just in terms of
exterior and Lie derivatives:
\begin{equation}\label{universalD} 2 \D_a f_b = 2 \D_{[a} f_{b]} + {\cal
L}_{\hat{f}}\, q_{ab} \end{equation}
where the vector field $\hat{f}^a = m^a m^b f_b$ is independent of
the choice of the unit space-like vector $m^a$ tangential to
$\Delta$. Thus, the action of $\D$ on these 1-forms is determined
by $q_{ab}$. Therefore, $\D$ is completely determined by its
action $\D_an_b =:S_{ab}$ on 1-forms $n$ satisfying $n\cdot \l =
-1$. Without loss of generality, we can assume that $n$ satisfies,
in addition,
\begin{equation} \label{n} {\cal L}_\l\, n \= 0\quad  {\rm and}\quad  dn \=
0\end{equation}
on $\Delta$. Then $S_{ab}$ is symmetric. Since $dn \= 0$, we have
$n = -dv$ for some function $v$ on $\Delta$ satisfying ${\cal
L}_\l v = 1$. The $v={\rm const}$ cross-sections will be assumed
to be topologically $S^1$ and denoted $\tilde{\Delta}$.

Now,
\begin{equation}\label{sl} \l^b S_{ab} \= - n_b \D_a \l^b \= -n_b
\grad_{\pback{a}} \l^b \= \omega_a \end{equation}
Thus, part of the `new' information in $\D$ is contained in the
1-form $\omega$ of Section \ref{s2.1}. The rest is contained in
the projection $\tilde\mu$ of $S_{ab}$ on $\tilde{\Delta}$:
$\tilde\mu:\= \tilde{m}^a \tilde{m}^b \D_a n_b$ where
$\tilde{m}^a$ is the unit vector field tangential to
$\tilde\Delta$. This function $\tilde\mu$ is the `transversal
expansion' of $n$ (see Appendix A).

Following the terminology used in higher dimensions \cite{abl1},
we will refer to the pair $(q, \D)$ as the \textit{intrinsic
geometry} of $\Delta$. Thus, the intrinsic geometry is determined
by a triplet $(m^a,\omega_a, \tilde\mu)$ on $\Delta$ for any
choice of $n$ satisfying (\ref{n}).

\subsection{Field equations and `free data' on a weakly\\ isolated
horizon} \label{s6.2}

Consider a weakly isolated horizon $(\Delta, [\l])$. In this
sub-section we will analyze the restrictions imposed by field
equations on the intrinsic geometry of $\Delta$ and extract the
free-data that suffices to determine this geometry.

We already know that the pair $(q,\D)$ satisfies
\begin{equation} \label{con} q_{ab} \l^b \= 0; \quad {\cal L}_\l\, q_{ab} \=
0;\quad \D_a q_{bc} \= 0; \quad \D_a\l^b \= \omega_a \l^b;\quad
{\cal L}_{\l}\, \omega \= 0; \end{equation}
and Equations (\ref{universalD}) and (\ref{sl}). We now want to
analyze the further constraints imposed by the full field
equations: $E_{ab} := R_{ab} - \frac{1}{2}R g_{ab} + \Lambda
g_{ab} - T_{ab} \=0$. We already saw in Section \ref{s2} that weak
isolation implies  $R_{ab}\l^a\l^b \=0$ and $R_{ab} \l^a m^b\= 0$.
Hence these projections of the field equations do not further
constrain the horizon geometry; they only restrict the matter
fields at the horizon. It turns out that the projections
$E_{ab}n^a\l^b\= 0$ and $E_{ab}n^a m^b \= 0$ dictate the
propagation of $\omega$ (or, the Newman-Penrose spin coefficients
$\epsilon$ and $\pi$ of Appendix A) off $\Delta$ while $E_{ab} n^a
n^b \= 0$ dictates the propagation of $\tilde\mu$ off $\Delta$.
Thus, these equations do not constrain the intrinsic horizon
geometry in any way. (For details, see Appendix A.1.)

The only new constraint comes from the equation $E_{ab}\tilde{m}^a
\tilde{m}^b \= 0$. Had $\Delta$ been a space-like surface, the
analogous equations would have given the evolution equations. In
the present case they also dictate an `evolution'
---that of $\tilde\mu$--- but now \textit{within} $\Delta$. We have:
\begin{equation}\label{con2} {\cal L}_\l \, \tilde\mu \= - \kl \tilde\mu +
\tilde{m}^a \tilde{m}^b \D_{(a} \tilde{\omega}_{b)} +
(\tilde{m}^a\tilde\omega_a )^2 + \frac{1}{2}\tilde{m}^a
\tilde{m}^b t_{ab}. \end{equation}
where $\tilde\omega_a = \tilde{m}_a \tilde{m}^b \omega_b$ is the
projection of $\omega$ on $\tilde\Delta$ and $t_{ab} = (T_{ab} +
(2\Lambda - T)g_{ab})$. This exhausts the field equations.

We can now specify the freely specifiable part of the horizon
geometry. Fix a 2-manifold $\Delta$, topologically $S^1\times
\Rbar$, and equip it with a vector field $\l$ along the $\Rbar$
direction. Fix a foliation by circles labeled by $v= {\rm const}$,
where ${\cal L}_\l\, v = 1$. On any one cross-section,
$\tilde\Delta$, fix a function $\tilde\mu$ and 1-forms $m$ and
$\omega$ such that $m$ is nowhere vanishing, $m\cdot \l = 0$, and
$\omega \cdot \l = \kl$, where $\kl$ is a constant. This is the
free data. `Evolve' it to all of $\Delta$ through ${\cal L}_\l\, m
= 0$, ${\cal L}_\l\, \omega = 0$ and (\ref{con2}), for a given
$t_{ab}$ on $\Delta$. Then the triplet $(m, \omega, \tilde\mu)$ on
$\Delta$ provides us with the intrinsic geometry of a weakly
isolated horizon.

Finally, under the mild assumption, ${\cal L}_\l\, R_{\pback{ab}}
\= 0$, we can integrate (\ref{con2}) to obtain:
\begin{equation} \label{mu1} \tilde\mu = e^{-\kl v} \tilde\mu^0 +
\frac{1}{\kl}\left[ \tilde{m}^a \tilde{m}^b
\D_{(a}\tilde{\omega}_{b)}+ (\tilde{m}^a\tilde\omega_a)^2 +
\frac{1}{2}\tilde{m}^a \tilde{m}^b t_{ab}\right] \end{equation}
if $\kl {\not\!\!\=} 0$ and
\begin{equation} \label{mu2} \tilde\mu = \tilde\mu^0 + \left[ \tilde{m}^a
\tilde{m}^b \D_{(a}\tilde{\omega}_{b)}+
(\tilde{m}^a\tilde\omega_a)^2 + \frac{1}{2} \tilde{m}^a
\tilde{m}^b t_{ab}\right]v \end{equation}
if $\kl \= 0$, where ${\cal L}_\l \,\tilde\mu^0 \= 0$. These
solutions bring out the generalization entailed in considering
weakly isolated horizons in place of Killing horizons: now, even
the \textit{intrinsic geometry} on the horizon (as defined above)
can be time-dependent. In spite of this, the zeroth and first laws
hold on any weakly isolated horizon.

\subsection{Good cuts of non-extremal weakly isolated
horizon}
\label{s6.3}

As in higher dimensions \cite{abl1}, every non-extremal weakly
isolated horizon admits a natural foliation. However, because the
Weyl tensor vanishes in 3 dimensions, and the first homology of
$\Delta$ is now non-trivial, the construction is now somewhat
different.

Recall first that every non-expanding horizon $\Delta$ carries a
natural closed 1-form $\underline{m}_a$. Since $d \underline{m} \=
0$, $\underline{m}$ generates a class in $H^1(\Delta)$, the first
cohomology of $\Delta$. Since $\Delta$ is isomorphic to $S^1\times
\Rbar$, we have:
\begin{equation}\label{eqn:cohomdelta}
  H^1(\Delta) = H^0(S^1) = \Rbar.
\end{equation}
Since the integral of $\underline{m}$ over any  cross section
yields $- 2\pi R_\Delta$,  $\underline{m}$ is not in the zero
class of $H^1(\Delta)$. Next, recall that the 1-form $\omega$ on
$\Delta$ is also closed. Therefore, it must be of the form
\begin{equation}\label{eqn:cohomarg}
  \omega_a = C \underline{m}_a + \partial_a\psi,
\end{equation}
for some constant $C\in\Rbar$ and some function $\psi$ on
$\Delta$. The function $\psi$ can now be used to define a
preferred foliation of the horizon. Since
\begin{equation}\label{eqn:lPsikappa}
\Lie_{\l}\, \psi = \l^a \omega_a = \kappa_{(\l)} \, ,
\end{equation}
with $\kappa_{(\l)}$ constant on the horizon, the lines $\psi =
{\rm const.}$ define a foliation of $\Delta$ provided
$\kappa_{(\l)}$ is non-zero. If the vector fields $[\l ]$ are
complete, as for example in stationary black holes, the leaves of
the foliation are guaranteed to be topologically $S^1$.

In the Newman-Penrose type framework of Appendix A, the preferred
foliation is characterized by the fact that the spin-coefficient
$\pi$ is constant on each leaf. Therefore, if the underlying
space-time is axi-symmetric in a neighborhood of $\Delta$, the
foliation coincides with the integral curves of the rotational
Killing vector. In BTZ space-times, $\omega$ is given by
(\ref{eqn:expromega}), $d\psi \= - \kl n = \kl dv$, whence $\psi
\= \kl v$, where $v$ is the Eddington-Finkelstein-like coordinate
(see (\ref{eqn:defefc}).

\subsection{Isolated horizons and uniqueness of \boldmath$[\l]$}
\label{s6.4}

Let $\Delta$ be a non-expanding horizon. Fix any cross-section,
choose any null normal to $\Delta$ on the cross-section and
propagate it by the geodesic equation to obtain a null normal
$\l_0$ on $\Delta$. Then $(\Delta, [\l_0])$ is an extremal weakly
isolated horizon. Denote by $v_0$ its affine parameter. Set
\begin{equation} \label{WIHall} \ell^a \= \kl\, (v_0 - B)\, \l_0^a
\end{equation}
where $\kl$ is a non-zero constant and  ${\cal L}_{\l} B \= 0$. It
is straightforward to check that $\l$ is a null normal with
surface gravity $\kl$ and every null normal with surface gravity
$\kl$ arises in this way. Similarly, any null normal with zero
surface gravity is given by
\begin{equation} \label{aff} \ell^{\prime a} \= \left(\frac{1}{A}\right)
\ell^a_{0}
\end{equation}
for some  function $A$ satisfying ${\cal L}_{\ell_{0}}\, A \= 0$.
To summarize, simply by restricting the null normals $\l$ to lie
in a suitably chosen equivalence class $[\l]$, from any given
non-expanding horizon $\Delta$, we can construct a weakly isolated
horizon  $(\Delta, [\l])$ which is either extremal or
non-extremal. However, because of the arbitrary functions involved
in (\ref{aff}) and (\ref{WIHall}), \textit{there is an infinite
dimensional freedom in this construction.}

It is natural to ask if this freedom can be reduced by
strengthening the notion of isolation. The answer is in the
affirmative.

{\bf Definition 3} An \textit{isolated} horizon $(\Delta, [\l])$
consists of a non-expanding horizon $\Delta$ equipped with an
equivalence class $[\l]$ of null normals satisfying
\begin{equation} \label{ih} (\D_a \,{\cal L}_\l - {\cal L}_\l\, \D_a ) X^b
\= 0 \end{equation}
for all vector fields $X$ tangential to $\Delta$. As before, $\l$
is equivalent to $\l^\prime$ if and only if $\l^\prime = c \l$ for
some positive constant $c$ and if condition (\ref{ih}) holds for
one null normal $\l$, it holds for all  null normals in $[\l]$. If
a non-expanding horizon $\Delta$ admits a normal $\l$ satisfying
(\ref{ih}), we will say its geometry \textit{admits an isolated
horizon structure}.

Before analyzing the remaining freedom in the choice of $[\l ]$,
let us examine the difference between weakly isolated and isolated
horizons. Note first that the weak isolation condition can be
written as
$$(\D_a\, {\cal L}_\l - {\cal L}_\l\, \D_a ) \l^b  \= 0 \ .$$
Thus, the present strengthening of that notion asks that the
commutator of $\D$ and ${\cal L}_\l$ vanish on \textit{all} vector
fields on $\Delta$, not just on $\l$.  Since the information in
$\D$ (beyond $q_{ab}$) is contained in the pair $(\omega,
\tilde\mu)$, the additional condition is \textit{precisely} ${\cal
L}_\l \, \tilde\mu \=0$. (While $\tilde\mu$ depends on the choice
of cross sections $\tilde\Delta$,\, $({\cal L}_\l\, \tilde\mu)$
does not.) Next, it is straightforward to check that, on any
isolated horizon, the pull-back of the full space-time curvature
is time-independent: ${\cal L}_\l\, R_{\pback{ab}} \= 0$. (Since
$R_{\pback ab}\l^b \=0$, it follows that $R_{\pback{ab}}$ is
Lie-dragged by \textit{every} null normal, $f\l^a$, to $\Delta$.)
Thus, on an isolated horizon, the restriction that led us to the
solution (\ref{mu1}) and (\ref{mu2}) for $\tilde\mu$ is
automatically satisfied. Therefore, in the non-extremal case,
$\tilde\mu^0$ of (\ref{mu1}) vanishes while in the extremal case
the quantity in the square brackets in (\ref{mu2}) must vanish. In
both cases, the freely specifiable data of Section \ref{s6.2} is
restricted; $\tilde\omega$ and $\tilde\mu$ can not be specified
freely on a cross-section $\tilde\Delta$, but are constrained.

Finally, let us analyze the issue of existence and uniqueness of
$[\l]$. Let $\Delta$ be a non-expanding horizon. We can always
choose a null normal $\l$ such that $(\Delta, [\l ])$ is a
non-extremal, weakly isolated horizon. Let us further suppose that
$(\Delta, [\l ])$ is not already an isolated horizon, i.e., ${\cal
L}_{\l} \, \tilde\mu \not\!\!\=0$ and ask if we can find another
null normal $\l^\prime = f \l$ such that $(\Delta, [\l^\prime])$
is isolated. Now, using the definition of weak isolation, it is
straightforward to check:
\begin{equation} [{\cal L}_\l\, \, , \D]_{a} K_b \= C_{ab}{}^c K_c \quad {\rm
where} \quad C_{ab}{}^c \= - N q_{ab}\, \l^c \end{equation}
for any 1-form $K_b$ on $\Delta$. The function $N$ is given by $N
\= {\cal L}_\l\, \tilde\mu$. Under the rescaling $\l^\prime \= f
\l$, we have
\begin{equation} \label{nprime} (N^\prime - N)q_{ab} \= 2 \omega_{(a} \D_{b)}
f + \D_a \D_b f \end{equation}
By transvecting this equation with $\l^b$ we obtain
\begin{equation} \D_a( {\cal L}_\l\, f + \kl f ) \=0\end{equation}
which implies
\begin{equation} f \=  B e^{-\kl v} + \frac{\kappa_{(\l^\prime)}}{\kl}, \quad
{\rm with}\,\,\, {\cal L}_\l\, B \= 0. \end{equation}
Thus, the key question now is: Does there exist a function $B$
such that $N^\prime \= 0$?   Substituting for $f$ in
(\ref{nprime}) and using the expression (\ref{mu1}) of $\mu$, we
conclude that $N^\prime$ vanishes if and only if B satisfies
\begin{equation} \label{M} {\bf M}\cdot B := [\tilde{\D}^2 + 2\tilde\pi
\tilde{\D} + \tilde\D \tilde\pi + \tilde\pi^2 + \tilde{m}^a
\tilde{m}^b R_{ab}] B \= \kappa_{(\l^\prime)} \tilde\mu^0
\end{equation}
on any cross-section $\tilde\Delta$ of $\Delta$, where
$\tilde{m}^a$ is the unit vector field tangential to
$\tilde\Delta$, $\tilde{\D} := \tilde{m}^a \D_a $ and $\tilde\pi
:= \tilde{m}^a \omega_a$. Note that, given any cross-section
$\tilde\Delta$, the operator is completely determined by the
non-expanding horizon geometry $(q_{ab}, \D )$.

We will say that the horizon geometry is \textit{generic} if the
operator ${\bf M}$ has trivial kernel. In this case, $B = {\bf
M}^{-1} (\kappa_{(\l^\prime)}\tilde{\mu}^0)$ is the unique
solution to (\ref{M}), where, without loss of generality, we have
assumed that (the $v$-dependence of) $f$ was so chosen that
$\kappa_{(\l^\prime)}$ is non-zero. Thus, every generic
non-expanding horizon admits a unique $[\l]$ such that $(\Delta,
[\l])$ is isolated horizon. Furthermore, this isolated horizon is
non-extremal.

What happens if the horizon geometry is non-generic? In this case,
Eq (\ref{M}) implies that if we choose $B$ to belong to the kernel
of ${\bf M}$, then $(\Delta, [\l^\prime])$ is an \textit{extremal}
isolated horizon. Thus, in contrast to the situation in higher
dimensions, \textit{every} non-expanding horizon admits an
isolated horizon structure. However, in the non-generic case,
uniqueness is not assured a priori; it may be possible to choose
another null normal $\l''$ such that $(\Delta, [\l''])$ is an
isolated horizon. However, assuming that $\Delta$ admits an
extremal isolated horizon structure and repeating the analysis
starting from (\ref{nprime}), it is easy to verify that: i)
$\Delta$ can not admit a distinct extremal isolated horizon
structure; and, ii) If it also admits a non-extremal isolated
horizon structure, then it admits a foliation on which
\textit{both} null normals ($\l^a$ and $n^a$) have zero
expansions. This is an extremely special situation.

To summarize, in contrast to higher dimensions, \textit{every}
non-expanding horizon admits an isolated horizon structure which
furthermore is unique except in extremely special cases.

\section{Discussion}
\label{s7}

In this paper, we introduced the notion of non-expanding, weakly
isolated and isolated horizons in 2+1-dimensional gravity
(Sections \ref{s2} and \ref{s6}), analyzed geometry of these
horizons (Sections \ref{s2} and \ref{s6}), and extended the zeroth
and first laws of black hole mechanics to weakly isolated horizons
(Sections \ref{s3}, \ref{s4} and \ref{s5}). The methods used were
the same as those employed in higher dimensions
\cite{afk,abl2,abl1} and the overall results are also analogous.
In particular, the first law again arises as a necessary and
sufficient condition for the evolution along a given space-time
vector field $t^a$ to preserve the symplectic structure in the
phase space, i.e., to be Hamiltonian. When they exist, the
Hamiltonians are given by a sum of two surface terms, one at
infinity and the other at the horizon. The term at infinity,
$E^t_\infty$,  represents the total energy of the system, while
the horizon term, $E^t_\Delta$, provides an expression of the
horizon energy, both defined by $t^a$. There is an infinite number
of vector fields providing Hamiltonian evolution, each with its
horizon energy and the corresponding first law. However, using our
knowledge of the stationary axi-symmetric black hole solutions in
2+1 dimensions, for each space-time in our phase space, we can
single out a preferred evolution field $t_o^a$ on $\Delta$ and
identify the corresponding horizon energy $E^{t_o}_\Delta$ as the
`horizon mass'. The corresponding first law is then the
`canonical' first law for mechanics of weakly isolated horizons.

There are, however, certain subtle but important differences from
higher dimensions. These arise because of: i) the peculiarities of
3 dimensional Riemannian geometry (particularly the fact that the
Weyl tensor vanishes identically); ii) the fact that the first
homology class of the horizon is now non-trivial (because the
topology of $\Delta$ is $S^1\times R$); and, more importantly,
iii) the boundary conditions \textit{at infinity}, which are
rather different from those in higher dimensions (especially the
ones satisfied by the electromagnetic potential). The first two of
these factors required us to modify our constructions and proofs
at several points in Sections \ref{s2} and \ref{s6}. The third
difference added a number of complications and twists in sections
\ref{s3}, \ref{s4} and \ref{s5}. We will conclude with a brief
discussion of an additional subtlety which is not discussed in the
main text.

In higher dimensions, it is natural to require that the
electromagnetic potential ${\bf A}$ go to zero at infinity, a
condition which freezes the asymptotic gauge freedom. In 2+1
dimensions, by contrast, since ${\bf A}$ diverges logarithmically
when the electric charge is non-zero, gauge freedom persists at
infinity. For concreteness and pedagogical simplicity, we chose to
fix it `by hand'  by specifying a precise asymptotic behavior of
${\bf A}$ (see Appendix B). Had we retained this freedom, all our
discussion would have gone through. The expressions of the
symplectic structure and angular momentum would have remained
unchanged. However, the electromagnetic scalar potential
$\Phi_{(t)} = -{\bf A}\cdot t$ would then have inherited an
ambiguity in the Cl\'{e}ment solution and this ambiguity would
have trickled down in the final expression of the horizon mass
$M_\Delta$ for general space-times. Thus, because the Maxwell
potentials diverge logarithmically, the horizon mass is in fact
ambiguous in presence of a non-zero electric charge.%
\footnote{As explained briefly in Appendix B, the allowed gauge
freedom ${\bf A} \mapsto {\bf A} +df$ is such that,
asymptotically, $df$ must be of the form $F(Q)dt$ for some
function $F$ only of electric charge. Therefore as in Section
\ref{s5.1}, in presence of a non-zero electric charge, there is a
freedom to add a function of charge to the scalar potential
$\Phi_{(t)}$. If we don't fix the gauge at infinity, this
ambiguity persists also in the Cl\'{e}ment solution and we are now
led to add an arbitrary function of charge to  the expression
(\ref{mass3}) of mass, subject only to the condition that this
function tend to zero in the limit of zero charge.}
In the main text, for simplicity of presentation, we eliminated
this ambiguity by hand through a specific choice of boundary
conditions on ${\bf A}$.

\section*{Acknowledgements} We would like to thank Christoffer
Beetle, Stephen Fairhurst, Badri Krishnan and Jerzy Lewandowski
for stimulating discussions. This work was supported in part by
the NSF grant PHY-0090091 and the Eberly research funds of Penn
State. JW was also supported through Duncan and Roberts
fellowships.

\appendix

\section{The 2+1 analog of the Newman-Penrose\\ formalism}
\label{sec:NP2plus1}

In this appendix we construct the 2+1 analog of the Newman-Penrose
(NP) formalism \cite{np}.

\subsection{Triads and spin-coefficients}

In place of the Newman-Penrose null tetrad, we will use a triad
consisting of two null vectors $\l^a$ and $n^a$ and a space-like
vector $m^a$, subject to:
    \begin{eqnarray}\label{eqn:lmntetrad}
        \l\cdot \l = n\cdot n & = & 0,\ \ m\cdot m = 1\\
        \l\cdot m & = & n \cdot m = 0\\
        \l\cdot n & = & -1.
    \end{eqnarray}
Note that, unlike in 3+1 dimensions, the vector $m^a$ is real.
Therefore, there will be no complex quantities appearing in our
2+1 analog of the NP formalism.

In terms of this triad, the space-time metric $g_{ab}$ can be
expressed as
\begin{equation}\label{eqn:metric}
  g_{ab} = -2 \l_{(a}n_{b)} + m_am_b\, ,
\end{equation}
and its inverse is given by
\begin{equation}\label{eqn:invmetric}
  g^{ab} = -2 \l^{(a}n^{b)} + m^am^b.
\end{equation}

We will now investigate spin-coefficients, i.e., the derivatives
of the triad vectors. The normalization and orthogonality
conditions on the triad vectors immediately lead to the following
relations:
\begin{eqnarray}
    \l^b\nabla_a \l_b = n^b\nabla_an_b & = & m^b\nabla_am_b  =0
    \label{eqn:tet1}\\
    \l^b\nabla_a m_b & = & - m^b\nabla_a \l_b\label{eqn:tet2}\\
    \l^b\nabla_a n_b & = & - n^b\nabla_a \l_b\label{eqn:tet3}\\
    n^b\nabla_am_b & = & - m^b\nabla_a n_b\label{eqn:tet4}
\end{eqnarray}
If we did not have any relations between $\l$, $n$ and $m$, we
would have had $3\times 3 \times 3 = 27$ independent spin
coefficients. However, the above equations impose $3\times 6 =18$
relations between them whence the number of independent parameters
is reduced to just $9$. To keep as close a contact with the
standard NP framework, our notation will closely follow that in
\cite{np}. However, since we have only a real spatial triad vector
$m^a$ rather than the pair $m^a, \bar{m}^a$ of the standard NP
framework, there are some inevitable discrepancies in factors of
$2$. The notation is summarized in tables \ref{tab:nl} --
\ref{tab:nm}.
\
\begin{table}
    \begin{center}
    \begin{tabular}{|cc||ccc|}
                \hline
                 & & & & \\
                 &       & $\l^b$ & $n^b$ & $m^b$ \\
                 \hline
                 \hline
                 & & \rule{1cm}{0pt}& \rule{1cm}{0pt}& \rule{1cm}{0pt}\\
        $D$      & $\l^a$ & 0 & $-\epsilon$ & $-\kappa_{\mbox{\scriptsize NP}}$ \\
                 & & & & \\
        $\Delta$ & $n^a$ & 0 & $-\gamma$ & $-\tau$ \\
                 & & & & \\
        $\delta$ & $m^a$ & 0 & $-\alpha$ & $-\rho$ \\
        \hline
    \end{tabular}
    \end{center}
    \caption{The components of $\nabla_a \l_b$.}
    \label{tab:nl}
\end{table}
\begin{table}
    \begin{center}
    \begin{tabular}{|cc||ccc|}
                \hline
                 & & & & \\
                 &       & $\l^b$ & $n^b$ & $m^b$ \\
                 \hline
                 \hline
                 & & \rule{1cm}{0pt}& \rule{1cm}{0pt}& \rule{1cm}{0pt}\\
        $D$      & $\l^a$ & $\epsilon$ & 0 & $\pi$ \\
                 & & & & \\
        $\Delta$ & $n^a$ & $\gamma$ & 0 & $\nu$ \\
                 & & & & \\
        $\delta$ & $m^a$ & $\alpha$ & 0 & $\mu$ \\
                 \hline
    \end{tabular}
    \end{center}
    \caption{The components of $\nabla_an_b$.}
    \label{tab:nn}
\end{table}
\begin{table}
    \begin{center}
    \begin{tabular}{|cc||ccc|}
                \hline
                 & & & & \\
                 &       & $\l^b$ & $n^b$ & $m^b$ \\
                 \hline
                 \hline
                 & & \rule{1cm}{0pt}& \rule{1cm}{0pt}& \rule{1cm}{0pt}\\
        $D$      & $\l^a$ & $\kappa_{\mbox{\scriptsize NP}}$ & $-\pi$ & 0 \\
                 & & & & \\
        $\Delta$ & $n^a$ & $\tau$ & $-\nu$ & 0 \\
                 & & & & \\
        $\delta$ & $m^a$ & $\rho$ & $-\mu$ & 0 \\
                 \hline
    \end{tabular}
    \end{center}
    \caption{The components of $\nabla_am_b$.}
    \label{tab:nm}
\end{table}

In terms of these spin coefficients, the covariant derivatives of
the triad vectors are given by:
\begin{eqnarray}
    \nabla_a\l_b & = & -\epsilon n_a\l_b +\kappa_{\mbox{\scriptsize NP}} n_am_b -\gamma
    \l_a\l_b\nonumber \\
    & & \ \ + \tau \l_a m_b + \alpha m_a\l_b -\rho
    m_am_b\label{eq:nablal}\\
    \nabla_an_b & = & \epsilon n_an_b -\pi n_am_b + \gamma
    \l_an_b\nonumber\\
    & & \ \ - \nu \l_am_b - \alpha m_an_b + \mu m_am_b\\
    \nabla_am_b & = & \kappa_{\mbox{\scriptsize NP}} n_an_b - \pi n_a\l_b + \tau
    \l_an_b\nonumber\\
    & & \ \  - \nu \l_a \l_b - \rho m_a n_b + \mu m_a \l_b
\end{eqnarray}
Hence the divergences of the triad vectors, used in the main text,
are given by:
\begin{eqnarray}
    \nabla_a\l^a & = & \epsilon - \rho\\
    \nabla_an^a & = & \mu - \gamma\\
    \nabla_am^a & = & \pi - \tau
\end{eqnarray}

We conclude this section with examples 2 (the generalized BTZ
black hole) and 3 (the Cl\'{e}ment solution) discussed in section
\ref{s2}. It is easy to verify that a desired triad in the
generalized BTZ space-time is given by:
\begin{eqnarray}
    \l^a & = & \partial_{v} +
    \frac{1}{2}(N)^2\partial_r - N^\phi\partial_\phi\\
    n^a & = & -\partial_r\\
    m^a & = & \frac{1}{r}\partial_\phi
\end{eqnarray}
The corresponding co-triads are
\begin{eqnarray}
    \l_a & = &  -\frac{1}{2}(N)^2 dv + dr\\
    n_a & = & -dv\\
    m_a & = & r N^\phi dv + r d\phi.
\end{eqnarray}
For this triad, the spin-coefficients are:
\begin{equation}
\begin{array}{rclrclrcl}
    \epsilon & = & \frac{f^\prime(r)}{2}-r(N^\phi)^2 & \gamma & = & 0 & \alpha & = & N^\phi \\
    & & & & & & & & \\
    \kappa_{\mbox{\scriptsize NP}} & = & 0 & \tau & = & N^\phi & \rho & = & -\frac{1}{2r}(N)^2 \\
    & & & & & & & & \\
    \pi & = & N^\phi  & \nu & = & 0 & \mu & = & -\frac{1}{r}
\end{array}
\end{equation}

For the BTZ black hole, the function $f(r)$ is given by
\begin{equation}
  f(r) = -\frac{M}{\pi} + \frac{r^2}{l^2}.
\end{equation}

For the Cl\'{e}ment solution, a convenient triad is
\begin{eqnarray}
    \l &=& -\frac{1}{2} N^{2} {\rm d}v + \frac{r}{K} {\rm d}r,
    \nonumber\\
    m &=& K {\rm d}\phi + K N^{\phi} {\rm d}v, \nonumber\\
    n &=& - {\rm d}v ,
\end{eqnarray}
and the corresponding spin-coefficients are given by:
\begin{eqnarray}
    \epsilon & = & -K \left( \Lambda + (N^{\phi})^{2} \right) -
    \frac{Q^{2}K}{4\pi r^{2}} \left( 1+ \omega N^{\phi} \right)^{2}
    \\
    \alpha & = & \pi = \tau = \frac{\omega Q^{2}}{4\pi K^{2}} \left( 1
- 2\ln
    \frac{r}{\bar{r_{0}}} \right) \\
    \rho & = & - \frac{N^{2}}{2K} \left( 1 +
    \frac{Q^{2}\omega^{2}}{4\pi r^{2}} \right) \\
    \mu & = & - \frac{1}{K} \left( 1 +
    \frac{Q^{2}\omega^{2}}{4\pi r^{2}} \right) \\
    \gamma & = & 0 \\
    \nu & = & \frac{Q^{2}\omega}{4\pi r^{2}} \\
    \kappa_{\mbox{\scriptsize NP}} & = & 0
\end{eqnarray}

\subsection{Curvature}

Since we are in $2 + 1 $ dimensions, all the information of th
curvature tensor is contained in the Ricci tensor $R_{ab}$. We
will thus calculate the different components of $R_{ab}$ in our
preferred triads. Our conventions for the Riemann tensor are:
\begin{equation}\label{eqn:relcurv}
    \nabla_a\nabla_bt^c - \nabla_b\nabla_at^c = -R_{abd}{}^c t^d.
\end{equation}
Using the tables of the previous section we can express components
of the Ricci tensor in terms of the spin coefficients as follows:
\begin{eqnarray}
    \Ricci \l \l & = & - \pi \,\kappa_{\mbox{\scriptsize NP}} +
  2\,\alpha \,\kappa_{\mbox{\scriptsize NP}}  -
  \epsilon \,\rho  -
  {{\rho }^2} + \kappa_{\mbox{\scriptsize NP}} \,\tau  +\nonumber\\
  & & \ \ \ +
   \Lie_{\l}\, \rho -  \Lie_m\, \kappa_{\mbox{\scriptsize NP}}\label{eqn:rll}\\
   \Ricci \l n & = & {{\pi }^2} - \pi \,\alpha  +
  2\,\gamma \,\epsilon  -
  \epsilon \,\mu  + \mu \,\rho  -
  \pi \,\tau  - \alpha \,\tau  +\nonumber\\
  & & \ \ \ +
  \Lie_\l\, \gamma  - \Lie_\l\, \mu  -
  \Lie_n \,\epsilon + \Lie_m \,\pi\\
  \Ricci \l m & = & 2\,\gamma \,\kappa_{\mbox{\scriptsize NP}}  - \pi \,\rho  -
  \rho \,\tau  + \Lie_\l \, \tau  -
  \Lie_n\, \kappa_{\mbox{\scriptsize NP}}\\
  \Ricci n \l & = & - \pi \,\alpha  +
  2\,\gamma \,\epsilon  -
  \gamma \,\rho  + \mu \,\rho  -
  \pi \,\tau  - \alpha \,\tau  +
  {{\tau }^2} +\nonumber\\
  & & \ \ \ +
  \Lie_\l\, \gamma -
  \Lie_n\, \epsilon + \Lie_n \,\rho -
  \Lie_m\, \tau \\
  \Ricci n n & = & - \gamma \,\mu   -
  {{\mu }^2} + \pi \,\nu  +
  2\,\alpha \,\nu  - \nu \,\tau  -
  \Lie_n \, \mu  + \Lie_m\, \nu \\
  \Ricci n m & = & - \pi \,\mu   +
  2\,\epsilon \,\nu  - \mu \,\tau  +
  \Lie_\l\, \nu - \Lie_n\, \pi\\
  \Ricci m \l & = & - \pi \,\epsilon  +
  \alpha \,\epsilon  +
  \gamma \,\kappa_{\mbox{\scriptsize NP}}  + \kappa_{\mbox{\scriptsize NP}} \,\mu  -
  \pi \,\rho  - \alpha \,\rho  +\nonumber \\
  & & \ \ \ +
  \Lie_\l\, \alpha - \Lie_m\, \epsilon \\
  \Ricci m n & = & \alpha \,\gamma  - \alpha \,\mu  +
  \epsilon \,\nu  + \nu \,\rho  -
  \gamma \,\tau  - \mu \,\tau  -\nonumber\\
  & & \ \ \ -
  {\Lie_n}\, \alpha + \Lie_m \, \gamma\\
  \Ricci m m & = & -{{\pi }^2} + \epsilon \,\mu  +
  2\,\kappa_{\mbox{\scriptsize NP}} \,\nu  +
  \gamma \,\rho  - 2\,\mu \,\rho  -
  {{\tau }^2} + \nonumber\\
  & & \ \ \ + \Lie_\l\, \mu -
  \Lie_n\, \rho - \Lie_m \,\pi +
  \Lie_m \,\tau
\end{eqnarray}
Finally, since the Ricci tensor is symmetric, we obtain the
following restrictions on the spin coefficients:
\begin{eqnarray}
    0 & = & {{\pi }^2} - \epsilon \,\mu  +
        \gamma \,\rho  - {{\tau }^2} -
        \Lie_\l\, \mu - \Lie_n\, \rho +
        \Lie_m\, \pi + \Lie_m\, \tau\\
    0 & = & \pi \,\epsilon  -
  \alpha \,\epsilon  +
  \gamma \,\kappa_{\mbox{\scriptsize NP}}  - \kappa_{\mbox{\scriptsize NP}} \,\mu  +
  \alpha \,\rho  - \rho \,\tau  -\nonumber\\
  & & \ \ \ -
  \Lie_\l\, \alpha + \Lie_\l\, \tau  -
  \Lie_n\, \kappa_{\mbox{\scriptsize NP}} + \Lie_m \,\epsilon\\
  0 & = & - \alpha \,\gamma  -
  \pi \,\mu  + \alpha \,\mu  +
  \epsilon \,\nu  - \nu \,\rho  +
  \gamma \,\tau  + \nonumber\\
  & & \ \ \ +  \Lie_\l\, \nu  -
  \Lie_n\, \pi + \Lie_n\, \alpha  -
  \Lie_m \, \gamma
\end{eqnarray}

\subsection{Triad rotations}
\label{sec:trans}

In this section we investigate how our spin-coefficients change
under Lorentz transformations. We begin with a boost in the plane
spanned by $\l^a$ and $n^a$:
\begin{eqnarray}
    \l^a & \longrightarrow & c\ \l^a\\
    n^a & \longrightarrow & \frac{1}{c}\ n^a\\
    m^a & \longrightarrow & m^a
\end{eqnarray}
Under the action of this boost, we have:
\begin{equation}
\begin{array}{rclrclrcl}
    \kappa^\prime_{NP} & = & c^2 \kappa_{\mbox{\scriptsize NP}} & \pi^\prime & = &
    \pi&\epsilon^\prime & = &  c \epsilon + \l^a\nabla_a c \\
     & & & & & & & &\\
    \tau^\prime & = & \tau & \nu^\prime & = & \frac{1}{c^2} \nu &
    \gamma^\prime & = & \frac{1}{c}\left(\gamma +\frac{1}{c}n^a\nabla_a
c\right) \\
     & & & & & & & & \\
    \rho^\prime & = & c \rho & \mu^\prime & = & \frac{1}{c} \mu &
    \alpha^\prime & = & \alpha + \frac{1}{c}m^a \nabla_a c
\end{array}
\end{equation}

Next, let us consider a null rotation:
\begin{eqnarray}
    \l^a & \longrightarrow & \l^a\\
    n^a & \longrightarrow & \frac{1}{2}c^2\l^a + n^a + c m^a\\
    m^a & \longrightarrow & c \l^a + m^a
\end{eqnarray}
The coefficients now transform as follows:
\begin{eqnarray}
    \kappa^\prime_{NP} & = & \kappa_{\mbox{\scriptsize NP}}\\
    \tau^\prime & = & \tau + \frac{1}{2} c^2 \kappa_{\mbox{\scriptsize NP}} + c \rho\\
    \rho^\prime & = & \rho + c \kappa_{\mbox{\scriptsize NP}}
\end{eqnarray}
\begin{eqnarray}
    \pi^\prime & = & \pi + \frac{1}{2} c^2 \kappa_{\mbox{\scriptsize NP}} + c\epsilon
    + \l^a\nabla_a c\label{eqn:transpi}\\
    \nu^\prime & = & \nu + \frac{1}{2}c^3 \epsilon + \frac{1}{4}
    c^4 \kappa_{\mbox{\scriptsize NP}} + c \gamma + \frac{1}{2} c^2 \tau + c^2
    \alpha + c^3 \rho + \frac{1}{2} c^2 \pi\nonumber \\
     & & \ \ \ \ \ + \frac{1}{2} c^2
    \l^a\nabla_a c + n^a \nabla_a c + c m^a\nabla_a c\\
    \mu^\prime & = & \mu + c^2\epsilon + \frac{1}{2} c^3
    \kappa_{\mbox{\scriptsize NP}} + c \alpha + c \pi + \frac{1}{2} c^2 \rho +\nonumber\\
     & & \ \ \ \ \ + c \l^a\nabla_a c + m^a \nabla_a c
\end{eqnarray}
\begin{eqnarray}
    \epsilon^\prime & = & \epsilon + c \kappa_{\mbox{\scriptsize NP}}\\
    \gamma^\prime & = & \gamma + \frac{1}{2} c^2 \epsilon +
    \frac{1}{2} c^3 \kappa_{\mbox{\scriptsize NP}} + c \tau + c \alpha + c^2 \rho\\
    \alpha^\prime & = & \alpha + c \epsilon + c^2 \kappa_{\mbox{\scriptsize NP}} + c
    \rho\label{transalpha}
\end{eqnarray}

\subsection{Components of the gravitational connection \boldmath$A$}
We can express the covariant derivative operator $\nabla_a$ in
terms of the connection 1-form $A_a^I$. Using the relation
\begin{equation}
  \nabla_a v_b = {A_a^I}_J\, v^J\,  e_{Ib},
\end{equation}
where $e_{Ib}$ is the triad, and using
\begin{equation}
  A_{a\ I}^{}J = \epsilon_{KI}{}^J A_a^K
\end{equation}
we arrive at the desired expression:
\begin{eqnarray}\label{eqn:anp}
    A_a^K  & = & (\pi n_a + \nu \l_a - \mu m_a) \l^K\nonumber\\
    & & \ \ \ \ + (\kappa_{\mbox{\scriptsize NP}} n_a + \tau \l_a - \rho m_a) n^K\nonumber\\
    & & \ \ \ \ \ \ \ \ + (-\epsilon n_a - \gamma \l_a +\alpha m_a )
    m^K
\end{eqnarray}
The analogous expression for the triad is just
\begin{equation}\label{eqn:forme}
    e^I_a = -\l_an^I - n_a \l^I + m_a m^I.
\end{equation}

\subsection{The Maxwell field and equations}
\label{sec:maxwellNP}
To conclude, let us consider the Maxwell field. The components of
the field strength ${\bf F}$ in our triad define the analogs of
the NP $\Phi_{i}$:
  \begin{equation}
   {\bf F} = \Phi_{0} n \wedge m + \Phi_{1} \l \wedge n +
   \Phi_{2} m \wedge \l.
  \end{equation}
Finally, the Maxwell equations are then given by
\begin{eqnarray}
   D\Phi_{1} - \delta\Phi_{0} & = & (\pi - \alpha)\Phi_{0} +
   \rho\Phi_{1} - \kappa_{\mbox{\scriptsize NP}} \Phi_{2}, \\
   2D\Phi_{2} - \delta\Phi_{1} & = & -\mu\Phi_{0} + 2\pi\Phi_{1} +
   (\rho - 2\epsilon)\Phi_{2}, \\
   2\Delta\Phi_{0} - \delta\Phi_{1} & = & (2\gamma - \mu)\Phi_{0} -
   2\tau\Phi_{1} + \rho\Phi_{2}, \\
   \Delta\Phi_{1} - \delta\Phi_{2} & = & \nu\Phi_{0} - \mu\Phi_{1} +
   (\alpha - \tau)\Phi_{2}.
\end{eqnarray}

\subsection{Horizons}

Because of the various boundary conditions, a number of
simplifications arise at the horizon $\Delta$. First, it is
convenient to assume that the null vector $n$ is exact, $dn \= 0$.
Then, $\alpha \= \pi$. If $\Delta$ is a non-expanding horizon, two
of the spin coefficients vanish; $\rho \= 0$ and $\kappa_{\mbox{\scriptsize NP}} \=
0$. Furthermore, $\l^a \D_a \pi \= m^a \D_a \epsilon$ and $\l^a
\D_a \tau \=0$. The Ricci tensor is constrained: $R_{ab}\l^a\l^b
\=0, R_{ab} \l^a m^b \= 0$. Finally, for the Maxwell field,
$\Phi_0 \= 0$ and $\l^a \D_a \Phi_1 \=0$.

On a weakly isolated horizon, spin coefficients are further
restricted: $\epsilon \= const$. In the non-extremal case,
$\epsilon \not\!\!\= 0$, the preferred foliation is characterized
by $\pi \= const$. On an isolated horizon, two further conditions
hold: $\l^a \D_a \mu \= 0$ and $\l^a \D_a (R_{cd} m^c m^d) \= 0$.

\section{Asymptotic behavior at spatial infinity}

In this Appendix we will specify the asymptotic fall-off of our
field variables. We will consider two cases: i) there are no
matter fields near infinity; and ii) the only matter field near
infinity is the Maxwell field. We separate these cases because, in
presence of charges, the second involves additional, significant
complications which do not arise in the first case. For both, we
will assume that in the neighborhood of spatial infinity
\begin{equation}
\begin{array}{rclrcl}
     A & \sim & \stackrel{\circ}{A} + \tilde{A}, &  e  & \sim & \stackrel{\circ}{e} + \tilde{e}, \\
       & & & & & \\
     {\bf A} & \sim & \stackrel{\circ}{{\bf A}} + \tilde{{\bf A}}, & \hstar {\bf F} & \sim & \stackrel{\circ}{\hstar{\bf F}} + \tilde{\hstar {\bf F}},
\end{array}
\end{equation}
where the quantities with the circle on top are certain background
fields (which we specify explicitly below) and the ones with tilde
are `smaller' quantities with specific fall-off (specified below)
in a radial coordinate $r$ defined by the background metric. Our
choice of asymptotic conditions is dictated by the following
stringent requirements: i) All explicitly known stationary black
hole solutions (that we are aware of) belong to the phase-space
defined by these conditions; ii) For fields satisfying these
asymptotic conditions, the action is finite (on- and off-shell)
and differentiable; iii) On the full phase, the Hamiltonian is
finite (on- and off-shell) and differentiable; iv) The symplectic
structure is well-defined; and, v) The boundary conditions are
preserved by the infinitesimal evolution.

\subsection*{Vacuum space-times}

In this case, the BTZ solutions naturally provide the required
background fields. Thus, we assume that a neighborhood of infinity
of every space-time of interest is diffeomorphic to a neighborhood
of infinity of the BTZ space-time. Then, in terms of the BTZ
coordinates $t,r,\phi$, we can specify the background co-triads
and connection:
\begin{eqnarray}
  \stackrel{\circ}{\l} & = & \frac{1}{2} \Lambda r^{2} {d}t +
  \frac{1}{2} {d}r, \\
  \stackrel{\circ}{n} & = & -{d}t + \frac{{d}r}{\Lambda r^{2}},
  \\
  \stackrel{\circ}{m} & = & r {d}\phi, \\
  \stackrel{\circ}{A}^{I} & = & {d}\phi \,\l^{I} - \frac{1}{2} \Lambda
r^{2}
  {d}\phi\,  n^{I}- \Lambda r {d}t \, m^{I} + \frac{1}{r} {d}r\, m^{I}
\end{eqnarray}
where $\l^I, n^I, m^I$ is a constant internal triad, satisfying
our orthogonality and normalization conditions with the fixed
internal metric $\eta_{IJ}$. An appropriate set of fall-off
conditions on the deviations $\tilde e$ and $\tilde A$ is given
by:
    \begin{equation}
    \begin{array}{rclrclrcl}
    & & & & & & & & \\
    \tilde{\l}_{t} &\sim &const. & \tilde{\l}_{r} &\sim& 1/r^{2}& \tilde{\l}_{\phi} & \sim & 1/r \\
    & & & & & & & & \\
    \tilde{n}_{t} & \sim & 1/r^{3} & \tilde{n}_{r} & \sim & 1/r^{4} & \tilde{n}_{\phi} & \sim & 1/r^{2} \\
    & & & & & & & & \\
    \tilde{m}_{t} &\sim & 1/r & \tilde{m}_{r}& \sim & 1/r^{3} & \tilde{m}_{\phi}& \sim &1/r^{2} \\
    & & & & & & & & \\
    l_{I} \tilde{A}^{I}_{t} & \sim & 1/r & l_{I} \tilde{A}^{I}_{r} %
    & \sim & 1/r^{2} & l_{I} \tilde{A}^{I}_{\phi}& \sim &const. \\
        & & & & & & & & \\
    n_{I} \tilde{A}^{I}_{t} & \sim & 1/r^{2}& %
    n_{I}\tilde{A}^{I}_{r}& \sim & 1/r^{3} & n_{I} \tilde{A}^{I}_{\phi} & \sim & 1/r^{3} \\
        & & & & & & & & \\
    m_{I} \tilde{A}^{I}_{t} & \sim &  1/r^{2} &
    m_{I}\tilde{A}^{I}_{r}& \sim & 1/r^{3} & m_{I} \tilde{A}^{I}_{\phi} & \sim &1/r
    \end{array}
   \end{equation}
where, in the Lagrangian framework, we consider only such histories for which variations of the constants fall-off as $1/r$ at infinity. 

\subsection{Electro-vacuum space-times}

To accommodate non-zero angular momentum \textit{and} charge, a
considerably more complicated choice of the background fields is
needed. A natural strategy would be to replace the BTZ background
with that provided by the Cl\'ement solution. Thus, for the
co-triad we are led to choose
\begin{eqnarray}
    \stackrel{\circ}{\l} & = & (\frac{1}{2} \Lambda r^{2} +
\frac{Q^{2}}{4\pi} \ln
    \frac{r}{\bar{r}_{0}}) {\rm d}t + \frac{1}{2} {\rm d}r, \\
    \stackrel{\circ}{n} & = & -{\rm d}t + (\Lambda r^{2})^{-1} {\rm
    d}r, \\
    \stackrel{\circ}{m} & = & -\frac{\omega Q^{2}}{2\pi r} \ln
    \frac{r}{\bar{r}_{0}} {\rm d}t +
    \frac{\omega Q^{2}\ln \frac{r}{\bar{r}_{0}}}{2\pi \Lambda r^{3}} {\rm
d}r
    + (r + \frac{\omega^{2}Q^{2}}{4\pi r}\ln
    \frac{r}{\bar{r}_{0}}) {\rm d}\phi, \\
    \stackrel{\circ}{A^{I}} & = & l^{I} \left( \frac{1}{8\pi} \Lambda \omega
    Q^{2} {\rm d}t + \frac{Q^{2}\omega}{8\pi r^{2}}{\rm d}r + (1 +
\frac{Q^{2}
    \omega^{2}}{4\pi r^{2}}){\rm d}\phi \right) + \nonumber \\
    & + & n^{I} \left( \frac{1}{4\pi} \Lambda \omega Q^{2} \ln
    \frac{r}{\bar{r}_{0}}{\rm d}t + \frac{Q^{2}\omega}{8\pi r^{2}} %
    {\rm d}r \right.- \nonumber\\
    & & \ \ \ - \left.(\frac{1}{2} \Lambda r^{2} + \frac{1}{8\pi}
    \Lambda Q^{2} \omega^{2} + \frac{1}{4\pi} Q^{2} \ln
    \frac{r}{\bar{r}_{0}}){\rm d}\phi \right) +  \\
    & + & m^{I} \left(( -r\Lambda - \frac{\Lambda Q^{2} \omega^{2}}{4\pi r}
    \ln \frac{r}{\bar{r}_{0}}) {\rm d}t + \frac{1}{r} {\rm d}r \right.+
    \nonumber\\
    & & \ \ \ \left. + \frac{\omega Q^{2}}{4\pi r}(1- 2 \ln \frac{r}{\bar{r}_{0}}) {\rm d}\phi%
    \right),\nonumber
\end{eqnarray}
and, for the Maxwell field,
\begin{eqnarray}
\stackrel{\circ}{{\bf A}} & = & Q \ln \frac{r}{\bar{l}} \left(
    {\rm d}t - \omega {\rm d}\phi \right) + \frac{1}{2} \Lambda Q
    \omega^{2} {\rm d}t, \\
    \hstar \, \stackrel{\circ}{\bf F} & = & Q \omega \Lambda {\rm
    d}t + Q {\rm d}\phi,
\end{eqnarray}
where, although the form of $\hstar \stackrel{\circ}{\bf F}$ is
determined by that of the background connection, we have displayed
it explicitly for convenience. The fall-off conditions on the
permissible deviations are given by
   \begin{equation}
    \begin{array}{rclrclrcl}
    & & & & & & & & \\
    \tilde{\bf A}_{t} & \sim & \frac{\ln r}{r} & \tilde{\bf A}_{r} & \sim & \frac{\ln r}{r^2} & \tilde{\bf A}_{\phi} & \sim & \frac{\ln r}{r} \\
    & & & & & & & & \\
    \hstar\tilde{\bf F}_{t} & \sim & \frac{1}{r} & \hstar\tilde{\bf F}_{r} & \sim & \frac{\ln r}{r^{2}} & \hstar\tilde{\bf F}_{\phi} & \sim & \frac{1}{r} \\
    & & & & & & & & \\
    \tilde{l}_{t} & \sim & \ln r/r & \tilde{l}_{r} & \sim & \ln r/r^{2} & \tilde{l}_{\phi} & \sim & 1/r \\
    & & & & & & & & \\
    \tilde{n}_{t} & \sim & 1/r^{3} & \tilde{n}_{r} & \sim & \ln r/r^{4} & \tilde{n}_{\phi} & \sim & \ln r/r^{2} \\
    & & & & & & & & \\
    \tilde{m}_{t} & \sim & \ln r/r^{2} & \tilde{m}_{r} & \sim & \ln r/r^{4} & \tilde{m}_{\phi} & \sim & \ln r/r^{2} \\
    & & & & & & & & \\
    l_{I} \tilde{A}^{I}_{t} & \sim & \ln r/r & l_{I}\tilde{A}^{I}_{r}&  \sim & \ln r/r^{3} & l_{I} \tilde{A}^{I}_{\phi} & \sim & 1/r \\
    & & & & & & & & \\
    n_{I} \tilde{A}^{I}_{t} & \sim & \ln r/r & n_{I}\tilde{A}^{I}_{r} & \sim & \ln r/r^{3} & n_{I} \tilde{A}^{I}_{\phi} & \sim & 1/r^{3} \\
    & & & & & & & & \\
    m_{I} \tilde{A}^{I}_{t} & \sim & \ln r/r^{2} & m_{I}\tilde{A}^{I}_{r}& \sim & \ln r/r^{3} & m_{I} \tilde{A}^{I}_{\phi}& \sim &\ln r/r^{2}\\
        & & & & & & & &
   \end{array}
   \end{equation}
In these conditions, the parameters $Q$, $\omega$ and
$\bar{r}_{0}$ do not depend on the coordinates $(t,r,\phi)$ and we
consider only such histories in the Lagrangian formulation for
which the variations of these parameters vanish at infinity at the
rate $1/r$.

We will conclude by pointing out a subtlety with respect to the
boundary condition on the electromagnetic vector potential ${\bf
A}$. In higher dimensions, one can simply require that ${\bf A}$
should vanish at spatial infinity. In 2+1 dimensions, by contrast,
if the electric charge is non-zero, ${\bf A}$ necessarily diverges
logarithmically. Now, the asymptotic form of ${\bf A}$ is not
fixed a priori by physical considerations; there is a possibility
of making a gauge transformation:
\begin{equation} {\bf A} \mapsto {\bf A} + d f \end{equation}
which can be non-trivial at infinity. For the Hamiltonian
framework of the main text to be well-defined, however, the
asymptotic form of $f$ is restricted; $d f$ must be of the form
$F(Q)dt$ at infinity. Note that the coefficient of $d t$ must be
constant in any given space-time and can only depend on the
charge; if it depended on other parameters ---$\omega$ and
$\bar{r}_0$--- then the term at infinity in (\ref{Xt2}) would not
be an exact variation and we would not be able to define energy at
infinity. Nonetheless, we do have a restricted gauge freedom. In
this Appendix, for concreteness and pedagogical simplicity, we
simply eliminated it `by hand' through our boundary conditions at
infinity. Had we retained this freedom, all our discussion would
have gone through. The expressions of the symplectic structure and
angular momentum would have remained unchanged. However, the
scalar potential $\Phi_{(t)}$ would then have inherited an
ambiguity in the Cl\'{e}ment solution through the undetermined
$F(Q)$ and this ambiguity would have trickled down in the final
expression of the horizon mass $M_\Delta$ for general space-times.
Thus, because the Maxwell potentials diverge logarithmically, a
priori the horizon mass is ambiguous in presence of a non-zero
electric charge. In the main text, for simplicity, we chose to
eliminate this ambiguity by hand through a specific choice of
boundary conditions, i.e., by fixing the Maxwell gauge at
infinity.

\end{document}